%% file: main.tex
\documentclass[floats,floatfix,showpacs,amssymb,prd,twocolumn,superscriptaddress,nofootinbib,nolongbibliography,reprint,prd,aps]{revtex4-2}
\usepackage{textcomp}
\usepackage{amssymb,amsmath,verbatim,mathtools,needspace,enumitem,etoolbox,graphicx,microtype,afterpage,bigints,gensymb,tabularx,xspace,siunitx}
\usepackage{aasmacros}
\usepackage{graphicx}
\usepackage{dcolumn}
\usepackage{bm}
\usepackage{lipsum}
\usepackage[usenames, dvipsnames, table]{xcolor}
\usepackage{physics}
\usepackage{soul}
\usepackage[normalem]{ulem}

\usepackage{float}
\makeatletter
\let\newfloat\newfloat@ltx
\makeatother
\usepackage{algorithm}
\usepackage[noend]{algpseudocode}
\usepackage{amsmath}
\AtBeginDocument{\RenewCommandCopy\qty\SI}
\ExplSyntaxOn
\msg_redirect_name:nnn { siunitx } { physics-pkg } { none }
\ExplSyntaxOff
\DeclareSIUnit\solarmass{\ensuremath{M_{\odot}}}
\DeclareSIUnit\parsec{pc}
\DeclareSIUnit\lightspeed{$c$}
\DeclareSIUnit\year{yr}
\DeclareSIUnit\arcsecond{as}
\DeclareSIUnit\astronomicalunit{AU}
\DeclareSIUnit\clight{\ensuremath c}

\definecolor{linkcolor}{rgb}{0.0,0.3,0.5}
\usepackage[colorlinks = true,
            linkcolor = linkcolor,
            urlcolor  = linkcolor,
            citecolor = linkcolor,
            anchorcolor = linkcolor]{hyperref}
\usepackage[capitalize]{cleveref}
\Crefname{equation}{Eq.}{Eqs.}
\Crefname{figure}{Fig.}{Figs.}
\Crefname{tabular}{Tab.}{Tabs.}
\Crefname{section}{Sec.}{Secs.}
\Crefname{subsection}{Sec.}{Secs.}

\usepackage{xcolor}
\definecolor{TealBlue}{HTML}{0072B2}
\definecolor{VioletPurple}{HTML}{CC79A7}
\definecolor{EmeraldGreen}{HTML}{009E73}
\definecolor{GoldenYellow}{HTML}{F0E442}

\usepackage{array}

\usepackage{orcidlink}
\begin{document}
\sisetup{range-phrase=-, range-units=single}

\author{Lorenzo Valbusa Dall'Armi~\orcidlink{0000-0002-0412-8058}}
\email{lorenzo.valbusadallarmi@df.unipi.it}
\affiliation{Dipartimento di Fisica “E. Fermi”, Università di Pisa, I-56127 Pisa, Italy}
\affiliation{INFN, Sezione di Pisa, I-56127 Pisa, Italy}
\author{Gabriele~Astorino~\orcidlink{0009-0005-1111-6708}}
\email{gabriele.astorino@phd.unipi.it}
\affiliation{Dipartimento di Fisica “E. Fermi”, Università di Pisa, I-56127 Pisa, Italy}
\author{Riccardo~Buscicchio~\orcidlink{0000-0002-7387-6754}}
\affiliation{Gran Sasso Science Institute (GSSI), I-67100 L’Aquila, Italy}
\affiliation{INFN, Laboratori Nazionali del Gran Sasso, I-67100 Assergi, Italy}
\affiliation{Dipartimento di Fisica “G. Occhialini”, Università degli Studi di Milano-Bicocca, Piazza della Scienza 3, 20126 Milano, Italy}
\affiliation{Institute for Gravitational Wave Astronomy \& School of Physics and Astronomy, University of Birmingham, Birmingham, B15 2TT, UK}
\author{Angelo~Ricciardone~\orcidlink{0000-0002-5688-455X}}
\affiliation{Dipartimento di Fisica “E. Fermi”, Università di Pisa, I-56127 Pisa, Italy}
\affiliation{INFN, Sezione di Pisa, I-56127 Pisa, Italy}
\author{Walter~Del~Pozzo~\orcidlink{0000-0003-3978-2030}}
\affiliation{Dipartimento di Fisica “E. Fermi”, Università di Pisa, I-56127 Pisa, Italy}
\affiliation{INFN, Sezione di Pisa, I-56127 Pisa, Italy}

\title{Global Fit Inference for LISA with \texttt{samsara++}}

\date{\today}

\begin{abstract} 

The Laser Interferometer Space Antenna (LISA) will observe a superposition of gravitational-wave signals, requiring joint inference of an unknown number of sources, parameters, and instrumental noise. To disentangle these contributions, that overlap in both time and frequency, a Global Fit (GF) analysis is needed. We present \texttt{samsara++}, a C++/CUDA sampler designed to address this trans-dimensional GF problem. Building on the  continuous-time MCMC framework of \texttt{samsara}, the algorithm combines birth-death processes for adding and removing sources, with mutation moves for parameter estimation. Birth and death are selected through posterior-dependent rates. Shared-memory parallelism, incremental likelihood updates, and indexed storage reduce the computational and memory costs, while parallel tempering improves the exploration of multimodal posteriors. We assess the performance  using simulated one-year LISA datasets containing Galactic double white dwarf binaries and stationary Gaussian instrumental noise. Analyses of populations containing 143 and 866 binaries start from empty source  models, without preliminary searches or proposals tailored to the injected population. These tests demonstrate source reconstruction and catalog generation, highlighting, at the same time,  the importance of source-count, overlapping signals, and incomplete mixing on joint source and noise inference. Then we perform a separate stress test, initialized at the injected parameters, exploring a population of 16,435 binaries, corresponding to 131,480 source parameters. This run completes 15 million iterations in approximately 16 hours using about 3 {\rm GB} of RAM. All together, these benchmarks show the computational feasibility of birth-death sampling for large Galactic-binary populations, and provide  a foundation for extending \texttt{samsara++} toward a comprehensive LISA global fit.

\end{abstract}

\maketitle

\section{Introduction}  

Over the past decade, the detection of gravitational waves (GWs) has opened a completely new, unexplored, window for astrophysics, fundamental physics, and cosmology. The LIGO-Virgo-KAGRA (LVK) collaboration has detected nearly 400 transient signals from binary black hole (BBH), binary neutron star (BNS), and black hole-neutron star (BHNS) systems~\cite{LIGOScientific:2016aoc,LIGOScientific:2018mvr,LIGOScientific:2020kqk,KAGRA:2021vkt,LIGOScientific:2025slb} in the $[10,10^3]\, \rm Hz$ band. At much lower frequencies, Pulsar Timing Array (PTA) collaborations, including NANOGrav~\cite{NANOGrav:2023gor}, EPTA + InPTA~\cite{EPTA:2023fyk}, PPTA~\cite{Zic:2023gta}, and CPTA~\cite{Xu:2023wog}, have reported the first compelling evidence for a stochastic GW background (SGWB) in the $\rm nHz$ band. The Laser Interferometer Space Antenna (LISA)~\cite{LISA:2017pwj,LISA:2024hlh}, designed to observe GWs at frequencies between $10^{-4}$ and $10^{-1}$ $\rm Hz$, will help bridge the vast frequency gap between PTAs and ground-based interferometers. By probing this largely unexplored regime, LISA will provide a wealth of information complementary to that obtained from both existing approaches~\cite{LISA:2022yao,LISACosmologyWorkingGroup:2022jok,LISA:2022kgy}.

A crucial difference between LISA and current detectors is how much ``brighter'' LISA sky will be compared to current detectors. Thus, a large number of different types of astrophysical sources are expected to be present in the detector data at all times. LISA is thus a signal-dominated detector, with multiple sources louder than the instrumental noise in each data segment~\cite{LISA:PrePhaseA,MockLISADataChallengeTaskForce:2007iof,MockLISADataChallengeTaskForce:2009wir,Vallisneri:2008ye}. Among the variety of deterministic sources which could be resolved by LISA, there are systems of massive black hole binaries (MBHBHs)~\cite{Klein:2015hvg}, extreme mass ratio inspirals (EMRIs)~\cite{Babak:2017tow}, and compact binary systems of stellar mass in an earlier stage of evolution with respect to the ones detected by ground-based interferometers~\cite{Sesana:2016ljz,Moore:2019pke}. Population-synthesis studies predict that the most numerous sources in the LISA frequency band are $10^6-10^7$ GWs emitted by double white dwarf binary systems (DWD) in the Milky Way~\cite{Hils:1990vc,Bender:1997hs,Nelemans:2001hp}. Such Galactic binaries contribute to the full data stream in the form of an incoherent stochastic foreground below a few $\rm mHz$~\cite{Hils:1990vc,Bender:1997hs,Nissanke:2012eh,Karnesis:2021tsh,Cornish:2017vip,Robson:2018ifk,Buscicchio:2025zeb}, while $10^3-10^4$ individual binaries could be resolved, mainly at higher frequencies~\cite{Nelemans:2001hp,Timpano:2005gm,Crowder:2006eu,Ruiter:2007xx,Korol:2017qcx,Littenberg:2020bxy,10.1093/mnras/stac3686,Lackeos:2023eub,Katz:2024oqg}.

The overlap of these signals in both time and frequency poses a major data-analysis challenge for LISA, underscoring the need for a global fit (GF) that jointly models resolvable sources, stochastic foregrounds, and backgrounds~\cite{Cornish:2005qw,Crowder:2006eu}. Several approaches have been proposed in the literature to explore a trans-dimensional parameter space encompassing different source types, with the unknown number of sources treated as an inference parameter.
The LISA pipelines GLASS~\cite{Littenberg:2020bxy,Littenberg:2023xpl} and Erebor~\cite{Katz:2024oqg} make use of reversible-jump (RJ)~\cite{green_rj} in the MCMC to explore different dimensionalities of the parameter space, while the pipeline Gee-Moo~\cite{Deng:2025wgk} is based on the product-space technique~\cite{carlin_productspace,Hee:2015eba}. Alternative pipelines rely on the maximum likelihood estimate (MLE) and evolutionary algorithms to iteratively subtract individual sources~\cite{Bouffanais:2015sya,Zhang:2021htc,Gao:2023oen,Strub:2023zxl} or perform a GF at fixed dimensions~\cite{Strub:2024kbe}. Recently, machine learning (ML) has also been applied to the GF via simulation-based inference (SBI)~\cite{Delmond:2026wik,Spadaro:2026evb,Tay:2026frc}. 

In this work, we adopt an approach based on spatial birth-death processes (BDP)~\cite{Preston_1975} in a continuous-time (CT) MCMC framework~\cite{Stephens2000BayesianAO,mohammadi2020continuoustimebirthdeathmcmcbayesian}. At each step of the CT-MCMC chain, we propose, for each signal class, new states with a larger (\emph{birth}), smaller (\emph{death}), or unchanged (\emph{mutation}) number of sources with respect to the current state. We then associate to each proposed point a rate that depends on the variation of the posterior between the proposed and current states. These rates are chosen to satisfy detailed balance, in analogy with the Metropolis--Hastings acceptance ratio~\cite{metropolis1953equation,hastings1970monte}. The collection of all signals across all classes (the \emph{society)} evolves according to a Poisson process specified by these rates. The next state is selected according to these rates, enabling efficient exploration of the trans-dimensional parameter space. Further details and a robust Python implementation of a Bayesian CT-MCMC algorithm based on BDP are provided in~\cite{Astorino:2025ccl}. The sampler, \texttt{samsara}, has been validated in several test cases, showing excellent performance in sampling the posterior over both the number  of sources and their parameters across different source types. 

One of the main advantages of \texttt{samsara} is that the computation of the rates is an intrinsically parallel operation; thus, it is possible to speed up convergence both in terms of computing time and in number of steps (\emph{generations}) of the Markov chain by increasing the number of proposals at each iteration. Python offers several routes to parallelism, but in its standard implementation the Global Interpreter Lock (GIL) limits the parallel execution of threads that share memory. Process-based parallelism avoids this limitation at the cost of replicating the CT-MCMC state in each process, which becomes demanding for the large LISA dataset. These constraints can be mitigated, for instance through compiled extensions or with frameworks like JAX~\cite{jax2018github}, but, for this application, we opted for C++/CUDA, which gives direct control over shared-memory parallelism for not purely numerical work. The C++/CUDA version of \texttt{samsara}, \texttt{samsara++}, is specifically designed for the GF and uses an optimized indexing scheme to avoid redundant memory allocations for arrays whose size could change. Therefore, the rates are computed in parallel, sharing large arrays involved in the likelihood, such as the residuals and covariance. Furthermore, quasi-monochromatic signals for DWD are heterodyned within the frequency band where their contribution is non-negligible, and both the signal and the likelihood evaluations are sliced according to the source parameters -- an operation that is far more efficient in C++ than Python. In \texttt{samsara++}, we further exploit parallelism by simultaneously evolving DWD populations in independent frequency bands using birth-death and mutation processes, and by introducing a trans-dimensional implementation of parallel tempering (PT)~\cite{Wang_2005,Earl_2005}. These two sampling schemes help improve the mixing and convergence of the chains, while making full use of the available CPU and GPU resources, thereby maximizing the computational efficiency of our algorithm. Although C++/CUDA was chosen to maximize performance, the CT-MCMC framework and the algorithmic and data-structure strategies developed here are language-independent, opening the way to a future Python implementation.

In this work, we validate the performance of \texttt{samsara++} using three mock GF analyses of DWD signals and instrumental noise. These test cases show that our sampler can recover the injected number of sources starting from initial empty states, without using information from the injections or astrophysical expectations to inform the prior on the number of sources. We discuss, in particular, how PT helps identify high-SNR signals, avoiding long burn-in phases characterized by an excess of sources. We present the posterior distributions for the number and parameters of the sources, checking the consistency with the injected values. In postprocessing, we also generate astrophysical catalogs with associated detection probabilities per source, finding a one-to-one correspondence with the catalog injected. To conclude, we reconstruct from the posterior samples the total deterministic and stochastic signals, comparing them with the simulated data. To present the performance of \texttt{samsara++}, we discuss the rate of posterior samples produced by the algorithm for a given population.

This paper is organized as follows: in Section~\ref{sec:CTMCMC} we briefly review the general theory of CT-MCMC, while in Section~\ref{sec:Inference model} we describe the likelihood, signal and noise models used for the GF inference. In Section~\ref{sec:Parallel computations} we present the main code optimization strategies and sampling shortcuts adopted by \texttt{samsara++}. In Sections~\ref{sec:SingleBin},~\ref{sec:Results}, and~\ref{sec:Stress_test} we show and discuss the test cases of the inference.

\section{Continuous Time Markov Chain Monte Carlo}
\label{sec:CTMCMC}

The LISA data streams will contain a superposition of GW signals from many sources of different classes, together with instrumental noise. In the \texttt{samsara} formalism developed in~\cite{Astorino:2025ccl}, we refer to a single source of GWs as an \textit{individual}, while we identify as \textit{species} the possible classes of signals. In the LISA context, species may represent e.g., DWDs, MBHBs, EMRIs, SGWBs, or instrumental noise. A \textit{population} is a collection of individuals of the same species, and the \textit{society} is the collection of populations of different species. The society specifies all the information about the state of the CT-MCMC at a given step. In a Bayesian framework, GF inference aims to determine the joint posterior distribution of the society parameters $\Theta$. For each species $\alpha$ whose population size is unknown, the number of individuals is itself included in $\Theta$, alongside the parameters describing those individuals. The posterior distribution is given by the Bayes' theorem
\begin{equation}
    p(\Theta \mid d, \, H) = \frac{{\pi}(\Theta \mid H)\mathcal{L}(d \mid \Theta,H)}{\mathcal{Z}(d,H)}\, , 
    \label{eq:BayesTheorem}
\end{equation}
where $d$ are the data observed by LISA and $H$ is the model hypothesis for the signals we are considering, e.g., the waveform and noise models. The prior distribution $\pi$ encodes the previous knowledge of the distribution of $\Theta$ for the model $H$, the likelihood function $\mathcal{L}$ represents the probability density of the data given $\Theta$ and $H$. The evidence $\mathcal{Z}$ is the marginal likelihood and ensures that the posterior is normalized. The Bayesian problem is fully solved once the prior distribution $\pi(\Theta | H)$ is chosen and the likelihood function $\mathcal{L}(\Theta)$ is specified, so that the posterior distribution $p(\Theta | d, \, H)$ can be formally determined.

A computational challenge in Bayesian data analysis is sampling the posterior distribution. For fixed-dimensional parameter spaces, commonly used methods are MCMC, Nested Sampling (NS)~\cite{Skilling2004,Skilling2006}, and Sequential Monte Carlo (SMC)~\cite{gordon_smc,delmoral_smc}. The LISA Global Fit, however, combines high dimensionality with uncertainty in the number of sources, requiring methods capable of exploring models of different dimensions. In RJ-MCMC~\cite{green_rj}, adopted by the pipelines described in~\cite{Littenberg:2023xpl,Katz:2024oqg}, a transition to a space of different dimensionality is proposed with some probability and accepted or rejected through a generalized MH acceptance probability. The acceptance ratio includes the absolute Jacobian determinant of the invertible transformation between parameter spaces augmented to satisfy dimension matching. Product space methods~\cite{carlin_productspace,Hee:2015eba}, implemented in~\cite{Deng:2025wgk},  instead embed competing models in a common augmented parameter space and introduce a discrete model indicator specifying, in this context, the number of sources. Marginalizing over the remaining parameters yields the posterior distribution of the source count. Both RJ-MCMC and product-space methods can infer this distribution without separately computing the evidence for each source count.

In this paper, we use \texttt{samsara}, a BDP-based CT-MCMC sampler presented in~\cite{Astorino:2025ccl}. Within the CT-MCMC framework, $\Theta$ is represented as a set of parameters, describing the individuals in the populations of all species, 
\begin{equation}
    \Theta = \left\{\left\{\theta_{i,\alpha}\right\}_{i=1}^{N_{\rm pop,\alpha}}\right\}_{\alpha=1}^{N_{\rm sp}}\, , 
\end{equation}
where $N_{\rm pop,\alpha}$ is the number of individuals in the species $\alpha$ at a given step of the chain and $N_{\rm sp}$ is the number of species considered in the model $H$. In our notation, $\theta_{i,\alpha}$ is a vector of dimension $N_{\rm par,\alpha}$. At each step of the chain, for each species $\alpha$ and process $p$ (birth, death, or mutation), a set of $M_{p,\alpha}$ new points in the parameter space $\Theta_{j,p,\alpha}$, with $j$ running from 1 to $M_{p,\alpha}$, is proposed from a set of distributions $h_{j,p,\alpha}(\cdot|\Theta)$. A transition rate $R_{j,p,\alpha}$, which can be fixed deterministically or computed adaptively from the posterior, is associated with $\Theta_{j,p,\alpha}$. In our implementation, we fix the mutation rate to a constant value. In addition, for the births and mutations we propose only one new point per species, thus $M_{b,\alpha}=M_{m,\alpha}=1$. For the death move, the points $\Theta_{j,d,\alpha}$ are obtained by removing the $j$-th individual of the $\alpha$-th population from the society, with unit density $h_{j,d,\alpha}$. Birth and mutation moves modify only a single individual in the society at each step of the MCMC\footnote{More sophisticated moves involving more than one individual, such as merge/split proposals, will be considered in future work.}. In order to ensure detailed balance~\cite{Stephens2000BayesianAO,cappe,mohammadi2020continuoustimebirthdeathmcmcbayesian}, the adaptive birth and death rates are computed according to
\begin{equation}
    \begin{split}
        R_{b,\alpha} =& \min \left\{1, \frac{(N_{\rm pop, \alpha} + 1)}{h_b(\Theta_{b,\alpha}|\Theta)}\frac{p(\Theta_{b,\alpha} | d)}{p(\Theta | d)} \right\} \, ,\\
        R_{j, d,\alpha} =& \min \left\{1, \frac{h_b(\Theta_{j,d,\alpha}| \Theta)}{N_{\rm pop,\alpha}} \frac{p(\Theta_{j,d,\alpha} | d)}{p(\Theta | d)} \right\}\, ,
        \label{eq:birth_death_rates}
    \end{split}
\end{equation}
while for a fixed birth rate the death rates are computed according to 
\begin{equation}
    R_{j,d,\alpha} = R_{b,\alpha} \frac{h_b(\Theta_{j,d,\alpha}| \Theta)}{N_{\rm pop,\alpha}} \frac{p(\Theta_{j,d,\alpha} | d)}{p(\Theta | d)} \, .
    \label{eq:birth_death_rates_fixed}
\end{equation}
The next state in the chain is selected among the candidate states from a multinomial distribution determined by the rates. If a birth or a death is selected, the new point in the parameter space is always accepted, while if a mutation process is selected, the new point is accepted or rejected with the MH scheme. The validity of the detailed balance condition, guaranteed by the computation of adaptive rates and by the MH acceptance, combined with the ergodicity assumption, ensures that the CT Markov chain has stationary distribution the target posterior distribution. 

A key feature of \texttt{samsara} is that it proposes, for each species and process, candidate points whose rates depend on the value of the posterior at those points. This approach  is designed to improve the efficiency of adding and removing sources of different species. Moreover, the posterior evaluations and rate calculations in Eq.~\eqref{eq:birth_death_rates} can be parallelized across proposals, since each candidate is evaluated relative to the same current state. The proposals adopted in \texttt{samsara++} for the DWD species are illustrated in Figure~\ref{fig:proposals}. For a detailed description of the algorithm, we refer the reader to~\cite{Astorino:2025ccl}.

In the next section, we discuss in detail the GF test case studied in this work with \texttt{samsara}, specifying the likelihood function, the waveforms employed, and the noise model.

\section{Inference model}
\label{sec:Inference model}

\subsection{Likelihood}
\label{sec:likelihood}

In this work, we apply \texttt{samsara} to perform efficient trans-dimensional Bayesian inference over DWD populations and instrumental noise. These tests highlight the code performance on large datasets and high-dimensional parameter spaces. In this analysis, we work in the Fourier domain, ignoring non-stationary features of the PSD of the noise~\cite{Edwards:2020tlp}, due to gaps~\cite{Carre:2010ra,Baghi:2019eqo,Burke:2025bun}, changes in time of the acceleration noise~\cite{PhysRevD.99.122003,Purdue_2007}, or glitches~\cite{Robson:2018jly,Spadaro:2023muy,Muratore:2025knh,Gair:2025chu}. A possible solution is to do a time-frequency analysis~\cite{Cornish:2020odn,Cornish:2025awt,Johnson:2026rrn,Bandopadhyay:2025fyx} and will be considered in future work. Following~\cite{Deng:2025wgk}, we consider $T_{\rm obs}=1\, \rm yr$ for the LISA data stream, which corresponds to a frequency resolution $\Delta f=1/T_{\rm obs}$. We then choose a cadence $\Delta t = 24$ s, setting the minimum frequency to the resolution ($f_{\rm min}=\Delta f$) to exclude the zero-frequency bin, and the maximum frequency to the Nyquist ($f_{\rm max}=1/(2\Delta t)$). The total number of points in the data series is $N_t = T_{\rm obs}/\Delta t$, while the number of frequency bins used is $N_f \equiv (f_{\rm max}-f_{\rm min})/\Delta f$. 

As in~\cite{Deng:2025wgk}, we assume TDI-1.5 Michelson measurements~\cite{2004PhRvD..69b2001S,Cornish:2003tz,Shaddock:2003dj,Tinto:2003vj,Tinto:2004wu,Babak:2021mhe} and work in the AET basis, where the Power Spectral Density (PSD) of the noise is diagonal. The data in the $I$-th channel are denoted as $d_I(f)$, while the residuals at a given step of the MCMC can be written as 
\begin{equation}
    r_I(\Theta_{\rm d},f) = d_I(f)-\sum_{\alpha=1}^{N_{\rm sp}^{\rm d}}\sum_{i=1}^{N_{\rm pop,\alpha}}h_I(\theta_{i,\alpha},f) \, , 
    \label{eq:residuals_full}
\end{equation}
where $N_{\rm sp}^{\rm d}$ is the number of species whose individuals emit deterministic signals, e.g., DWD, MBHB, EMRI. Following the notation introduced in Section~\ref{sec:CTMCMC}, we have defined the parameters of the deterministic signals at a given step of the chain as 
\begin{equation}
    \Theta_{\rm d} \equiv \left\{\left\{\theta_{i,\alpha}\right\}_{i=1}^{N_{\rm pop,\alpha}}\right\}_{\alpha=1}^{N_{\rm sp}^{\rm d}}\, .
    \label{eq:Theta_det}
\end{equation}
In this analysis, we focus only on the deterministic signals emitted by DWDs, but our results could easily be generalized to the case of multiple deterministic species. We model the noise process as a zero-mean, stationary auto-covariance process, thus, according to the maximum entropy principle, the likelihood of the data is a multivariate Gaussian distribution
\begin{equation}
    \log\mathcal{L}_G(d|\Theta) = -\frac{1}{2}\sum_{f,I}\left[\log\left[2\pi S_I(f,\Theta_{\rm n})\right]+\frac{|r_I(f,\Theta_{\rm d})|^2}{S_I(f,\Theta_{\rm n})}\right] \, ,
    \label{eq:Whittle}
\end{equation}
with $I = \{A,E,T\}$ and $\Theta_{\rm n}$ the analogue of Eq.~\eqref{eq:Theta_det} for instrumental noise. The quantity $S_I$ is defined in terms of the PSD of the noise as
\begin{equation}
    S_I(f,\Theta_{\rm n}) \equiv \frac{{\rm PSD}(f,\Theta_{\rm n})}{4\Delta f}\, .
\end{equation}

Since the number of DWDs resolvable by LISA is expected to be around $10^4$~\cite{Nelemans:2001hp,Timpano:2005gm,Crowder:2006eu,Ruiter:2007xx,Korol:2017qcx,Littenberg:2020bxy,10.1093/mnras/stac3686,Lackeos:2023eub,Katz:2024oqg}, the computation of the full residuals in Eq.~\eqref{eq:residuals_full} would require approximately $N_{\rm DWD} N_{\rm ch} N_f$ operations. In addition, as described in Section~\ref{sec:CTMCMC}, \texttt{samsara} relies on the computation of birth, death, and mutation rates; therefore, $N_{\rm DWD}$ likelihood evaluations for the death rates are performed at each step of the MCMC. In the case of a full likelihood evaluation of $N_{\rm ch}N_f$ points, this is computationally prohibitive both in terms of memory and computational time. 

To overcome this issue, we adopt an approximation scheme for the computation of the rates of DWDs. First, we store the residuals and the log-likelihood $\log\mathcal{L}$ at each generation. Secondly, we notice that, as we will discuss in Section~\ref{sec:waveform},  
the signals emitted by DWDs are peaked in a frequency band $\mathfrak{f}$ much smaller than the LISA band. The $\mathfrak{f}$ band depends on the DWD parameters, but typically contains $N_{\rm het}(\theta)\approx 10^2-10^4$ points. Since for our birth, death, and mutation proposals only one source is added, removed, or mutated in the society, it is natural to compute the variation of the likelihood according to  
\begin{equation}
    \begin{split}
    \delta\log\mathcal{L}_G(d|\Theta_d^\prime,\Theta) =& \log\mathcal{L}_G(d|\Theta^\prime)-\log\mathcal{L}_G(d|\Theta)\\
    =& \ -\frac{1}{2}\sum_{f\in \mathfrak{f}, I} \frac{\left|\delta r_I\right|^2+\delta r_I^* r_I+\delta r_Ir_I^*}{S_I} \, ,
    \end{split}
    \label{eq:delta_Whittle}
\end{equation}
where $\Theta^\prime\equiv \{\Theta_d^\prime, \Theta_s\}$. In Eq.~\eqref{eq:delta_Whittle}, we also defined the variation of the residuals for the three processes allowed in \texttt{samsara} as 
\begin{equation}
    \delta r_I(\Theta_d^\prime, \Theta,f) = \begin{cases}
         -h(\theta_{\rm new},f) \hspace{6em} \rm birth \\
          h(\theta_{\rm old},f) \hspace{7.em} \rm death \\
          h(\theta_{\rm old},f) -h(\theta_{\rm new},f) \hspace{1.5em} \rm mutation 
    \end{cases}
    \label{eq:delta_residuals}
\end{equation}
where $\theta_{\rm new}$ is the new point in the DWD population, while $\theta_{\rm old}$ is the removed or mutated one. If the species selected for the Poisson evolution of \texttt{samsara} is DWD, we update the residuals and log-likelihood using 
\begin{equation}
    \begin{split}
        r_I(\Theta^\prime_d,f) =& \ r_I(f,\Theta_d,f)+\delta r_I( \Theta^\prime_d, \Theta,f) \, , \\
        \log \mathcal{L}_G(d|\Theta) =& \ \log \mathcal{L}_G(d|\omega,\sigma)+\delta \log\mathcal{L}_G (d|\Theta^\prime_d,\Theta)\, .       
    \end{split}
    \label{eq:update_residuals_likelihood}
\end{equation}
The computational cost of \texttt{samsara} for DWD is $\mathcal{O}(N_{\rm DWD}N_{\rm ch}N_{\rm het})$ and $\mathcal{O}(N_{\rm ch}N_{\rm het})$ operations for updating the rates and log-likelihood respectively, where $N_{\rm het}$ is the number of frequency points in $\mathfrak{f}$, defined in Eq.~\eqref{eq:N_het}.

To sample the PSD parameters, one could bin the frequencies into intervals much larger than the frequency resolution; see, e.g.,~\cite{Caprini:2019pxz,Buscicchio:2025zeb}. However, since noise parameters vary less frequently than DWD parameters, the evaluation of the likelihood over $N_{\rm ch}N_f$ points is not a computational bottleneck. In addition, the computation of the full likelihood, Eq.~\eqref{eq:Whittle}, has been parallelized over channels and frequencies on both CPU and GPU.

\subsection{Waveform}
\label{sec:waveform}

Galactic binaries emit GWs in the LISA band during the early stage of their inspiral and can be modeled as mildly chirping sinusoids. For this reason, following~\cite{Takahashi:2002ky,Krolak:2004xp,Ruiter:2007xx,Kremer:2017xrg,Toubiana:2024qxc}, for DWD we consider the parameters 
\begin{equation}
    \theta_{\rm DWD} \equiv \left\{f_{\rm GW}, \phi_0, A_0, \sin\beta, \lambda, \cos\iota, \psi,  \dot{f}_{\rm GW} \right\}\, , 
    \label{eq:dwd_parameters}
\end{equation}
where $f_{\rm GW}$ is the initial frequency, $\phi_0$ is the initial phase, $A_0$ is the time independent amplitude contribution, $\beta$ is the ecliptic latitude, $\lambda$ the ecliptic longitude, $\iota$ is the inclination angle between the angular momentum of the binary and the observer, $\psi$ is the polarization angle, and $\dot{f}_{\rm GW}$ is the frequency derivative\footnote{We assume here that the second derivative of the frequency is negligible.}. The GW phase of the chirping sinusoids is given by 
\begin{equation}
    \Phi_{\rm GW}(\theta_{\rm DWD}, t) \equiv \phi_0 +2\pi f_{\rm GW} t+\pi \dot{f}_{\rm GW}t^2 \, .
\end{equation}
At leading order, the amplitude depends on the chirp mass $M_c$ and luminosity distance $D_L$. Accounting for the slow frequency evolution to first order gives
\begin{equation}
    A = A_0\left(1+\frac{2}{3}\frac{\dot{f}_{\rm GW}}{f_{\rm GW}}t\right), \quad A_0 = \frac{2M_c G^{5/3}}{c^4 D_L}\left(\pi M_c f_{\rm GW}\right)^{2/3} \, , 
    \label{eq:amplitude_dwd}
\end{equation}
where the time-dependent factor is obtained from the perturbative expansion of the amplitude with respect to the slow change of the frequency of the DWD. The two GW polarizations are 
\begin{equation}
    \begin{split}
        \tilde{H}_+(\theta_{\rm DWD},t) =& A \frac{1+\cos^2(\iota)}{2}\cos\left[\Phi_{\rm GW}(\theta_{\rm DWD}, t)\right] \, , \\
        \tilde{H}_\times(\theta_{\rm DWD},t) =& A \cos\iota\,\sin\left[\Phi_{\rm GW}(\theta_{\rm DWD}, t)\right] \, . 
        \label{eq:DWD_polarizations}
    \end{split}
\end{equation}
The signal in LISA channel $I$ can be written schematically as
\begin{equation}
    H_I(\theta_{\rm DWD},t) = \sum_{p=+,\times}F_{p,I}(\theta_{\rm DWD},t)\tilde{H}_p(\theta_{\rm DWD},t) \, ,
    \label{eq:LISA_signal}
\end{equation}
where $F_{p,I}$ denotes the detector response, which depends on $\beta$, $\lambda$, and $\psi$. The quantity $h_I$ introduced in Eq.~\eqref{eq:residuals_full} is the Discrete Fourier Transform (DFT) of~\eqref{eq:LISA_signal}. Further details of the LISA response are given in~\cite{Cornish:2002rt,Cornish:2007if}.

Because the intrinsic frequency evolution is slow,
$\dot{f}_{\rm GW}T_{\rm obs}\ll f_{\rm GW}$, the Fourier-domain signal is concentrated in a narrow frequency interval near $f_{\rm GW}$.  Its spectral width reflects the intrinsic chirp and the modulation induced by the LISA response. Following~\cite{Hellings:2002qb,Cornish:2007if}, for mildly chirping binary sources, the bandwidth of the signal can be estimated by accounting for the finite observation time, the orbital Doppler modulation, and the intrinsic chirp as
\begin{equation}
    \Delta f_{\rm GW}(\theta_{\rm DWD}) = \frac{2}{T_{\rm obs}}\left(4+2\pi f_{\rm GW}\frac{R_{\rm ES}}{c}\sin\theta\right) +\dot{f}_{\rm GW}T_{\rm obs} \, ,
    \label{eq:deltaf_DWD}
\end{equation}
with $R_{\rm ES}=1\, \rm AU$ the radial distance of the guiding center of LISA. The frequency band $\mathfrak{f}$ over which we compute the variation of the likelihood, Eq.~\eqref{eq:delta_Whittle}, and residuals, Eq.~\eqref{eq:delta_residuals}, is defined as 
\begin{equation}
    \begin{split}
        \mathfrak{f} \equiv & [f^{\mathfrak{f}}_{\rm min}, f^{\mathfrak{f}}_{\rm max}] \, , \\
        f^{\mathfrak{f}}_{\rm min} \equiv& \min\left[\max\left(f_{\rm min}, f_{\rm GW}-5\Delta f_{\rm GW}\right), f_{\rm max}-\Delta f\right] \, , \\
        f^{\mathfrak{f}}_{\rm max} \equiv& \max\left[\min\left(f_{\rm max}, f_{\rm GW}+5\Delta f_{\rm GW}\right), f_{\rm min}+\Delta f\right] \, , 
        \label{eq:DWD_frequency_range}
    \end{split}
\end{equation}
with $f_{\rm min}$, $f_{\rm max}$ the minimum and maximum frequencies accessible by LISA, defined in Section~\ref{sec:likelihood}. The interval is chosen to retain the relevant signal power while remaining within the analysis band.

The computation of the waveform in the time domain requires evaluating Eq.~\eqref{eq:LISA_signal} at the $N_t$ points of the time series. However, since for a DWD we are interested only in the interval $\mathfrak{f}$, we heterodyne the signal~\cite{Cornish:2007if}. More specifically, we compute the waveform on a shorter time series with $N_t^{\mathfrak{f}}\ll N_t$ points and take its DFT after applying a phase shift. Since the DWD signal is confined in $\mathfrak{f}$, the DFT is identical, up to numerical precision, to the one obtained from the full time series. 
Heterodyning speeds up the computation of the waveforms by a factor $N_t/N_t^{\mathfrak{f}}$, while restricting the DWD signal to $\mathfrak{f}$ accelerates the computation of the likelihood by a factor $N_f/N_{\rm het}$, where the number of frequency points of the heterodyned DWD signal is defined as
\begin{equation}
    N_{\rm het}(\theta_{\rm DWD})\equiv\frac{f_{\rm max}^{\mathfrak{f}}-f_{\rm min}^{\mathfrak{f}}}{\Delta f} \, .
    \label{eq:N_het}
\end{equation}

In this work, we do not rely on publicly available codes, such as GBGPU~\cite{Cornish:2007if,Robson:2018svj,Katz:2022izt}; instead, we modify and upgrade the \texttt{BALROG}~\cite{2019PhRvD.100h4041B,Roebber:2020hso,Finch:2022prg,Moore:2023nbw} waveforms in \texttt{samsara++}, improving the heterodyning scheme and porting them to GPU with CUDA.

\subsection{LISA instrumental noise}
\label{sec:Instrumental noise}

For the instrumental noise of LISA, we consider both the test-mass acceleration and the Optical Metrology System (OMS) noises. The PSD of these two contributions can be written in terms of two amplitudes as~\cite{LISASciRD}
\begin{equation}
    \begin{split}
        S_{\rm pm}(f) =& \left[1+\left(\frac{0.4\, {\rm mHz}}{f}\right)^{2}\right]\,
        \left[1+\left(\frac{f}{8\,{\rm mHz}}\right)^{4}\right]\frac{S_{\rm acc}}{(2\pi c f)^2} \, , \\
        S_{\rm op}(f) =&  \left[1+\left(\frac{2\, \rm mHz}{f}\right)^4\right]\left(\frac{2\pi f}{c}\right)^2 S_{\rm oms} \, .
        \label{eq:acc_oms_psds}
    \end{split}
\end{equation}
Following~\cite{Deng:2025wgk}, we adopt the equal-arms approximation for the three LISA channels~\cite{Babak:2021mhe}, and identical acceleration and OMS noise PSDs across all links. In such basis, the noise PSDs in the $X$, $Y$, and $Z$ channels of LISA are identical; therefore, in the AET basis, the PSD of the instrumental noise is diagonal. Ref.~\cite{Prince:2002hp} showed that the acceleration and optical noise contribute to the PSD through the combinations
\begin{equation}
    \begin{split}
        N_{A,E}(f) = & 
        8\sin^2\left(\frac{f}{f_*}\right)\left[N_{A,E}^{\rm pm}(f)+N_{A,E}^{\rm op}(f)\right] \, , \\
        N_T(f) = &\,
        16\sin^2\left(\frac{f}{f_*}\right)
        \Biggl\{\left[1-\cos\left(\frac{f}{f_*}\right)\right]S_{\rm op}(f)
            \\
            &\hspace{6.5em}+ 8\sin^4\left(\frac{f}{2f_*}\right)S_{\rm pm}(f)
        \Biggl\} \,
    \end{split}
    \label{eq:LISA_psd}
\end{equation}
where we have defined the characteristic frequency of LISA $f_*\equiv c/(2\pi L)$, with $L=2.5\times 10^6\, \rm km$ the length of the arm. In Eq.~\eqref{eq:LISA_psd}, we defined for convenience the acceleration and OMS terms entering the \(A,E\) noise PSDs
\begin{equation}
    \begin{split}
        N_{A,E}^{\rm pm}(f) \equiv& \left[6+4\cos\left(\frac{f}{f_*}\right)+2\cos\left(\frac{2f}{f_*}\right)\right]S_{\rm pm}(f) \, , \\ 
        N_{A,E}^{\rm op}(f) \equiv& \left[2+\cos\left(\frac{f}{f_*}\right)\right] S_{\rm op}(f) \, . 
    \end{split}
\end{equation}
Since in our analysis we neglect data gaps, glitches, and sensitivity drift over time~\cite{LISA:2024hlh}, the noise parameters involved in our Bayesian analysis are just $S_{\rm acc}$ and $S_{\rm oms}$. For simplicity, we parametrize the two amplitudes in terms of the reference values
\begin{equation}
    \begin{split}
    S_{\rm acc} =& \left(3\times 10^{-15}\,{\rm m\,s^{-2}\,Hz^{-1/2}}\right)^2 \, r_{\rm acc} \, ,  \\
    S_{\rm oms} =& \left(15\times 10^{-12}\,{\rm m\,Hz^{-1/2}}\right)^2\, r_{\rm oms} \, . 
    \end{split}
    \label{eq:noise_amplitude_parametrization}
\end{equation}
Therefore, the noise parameters involved in the inference are $\theta_{\rm noise}\equiv\{r_{\rm acc},r_{\rm oms}\}$.

\subsection{Astrophysical population}
\label{sec:Astrophysical Model}

To characterize the Galactic DWD population, we specify a joint probability distribution for the source parameters introduced in Eq.~\eqref{eq:dwd_parameters}, together with a probability mass function for the total number of systems. These distributions are taken from~\cite{Korol:2020lpq,Buscicchio:2024wwm} and the  catalog used to generate the injected signals is similar to that used in the LDC-2a ``Sangria'' dataset~\cite{LDC_Sangria}.

The astrophysical catalog is obtained using the population synthesis code \texttt{SeBa}~\cite{Toonen_2012,PortegiesZwart:1996,Nelemans:2001hp,Toonen_2017}, which evolves binaries from the Zero-Age Main Sequence (ZAMS) until the DWD formation and is calibrated against recent DWD observations. \texttt{SeBa} provides the distribution of the white dwarf (WD) masses, the orbital separations at formation, and the DWD formation times. 

The stellar density distribution of the model consists of a central bulge and an extended disk. The bulge is spherically symmetric, while the disk has an exponential radial density profile and an isothermal vertical density profile~\cite{Nelemans:2003ha,Korol:2018wep}. 

The initial phase is uniform in $[0,2\pi]$, the polarization angle in $[0,\pi]$, and the cosine of the inclination angle in $[-1,1]$.

Once the catalog is generated, the frequency is connected to the orbital separation by Kepler's law and the amplitude is determined by Eq.~\eqref{eq:amplitude_dwd}. We then assume that binaries evolve as isolated systems and that interactions between individual components are negligible. Therefore, the frequency derivative is obtained by the GW-driven orbital decay, 
\begin{equation}
        \dot{f} = \frac{96}{5}\pi^{8/3}\left(\frac{GM_c}{c^3}\right)^{5/3}f^{11/3} \, .
    \label{eq:frequency_derivative}
\end{equation} 

The full catalog contains approximately $2\times 10^7$ sources, among which roughly $2\times 10^4$ have SNR larger than 8 for one year of observations with LISA.

\subsection{Priors on the parameters}
\label{sec:Priors}

We adopt DWD parameter priors that are independent of the population used to generate the injected signals. In particular, the GW frequency is uniformly distributed over $f_{\rm min}$ and $f_{\rm max}$ defined in Section~\ref{sec:likelihood}. We assign uniform priors to $\phi_0$ and  $\lambda$ over $[0, 2\pi]$, to $\cos \iota$ and $\sin \beta$  over $[-1,1]$ and  to $\psi$ over $[0, \pi]$. These angular priors correspond to isotropic sky positions and orbital orientations. To set the bounds on the frequency derivative, we use the leading-order GR relation for circular binaries evolving solely through GW emission, Eq.~\eqref{eq:frequency_derivative}. Given $f_{\rm min}$ and $f_{\rm max}$, it is possible to extract $\dot{f}_{\rm min}$ and $\dot{f}_{\rm max}$ by using the minimum and maximum chirp mass for a DWD system, computed assuming $m_1=m_2=0.15\, M_\odot$ and $m_1=m_2=2.5\, M_\odot$, respectively. We then adopt a uniform prior on $\dot{f}_{\rm GW}$ between $\dot{f}_{\rm min}$ and $\dot{f}_{\rm max}$. Finally, following~\cite{Deng:2025wgk}, we assign the GW amplitude a log-uniform prior over $10^{-24}$ and $10^{-20}$.  

We assign independent uniform priors over $[10^{-1}, 10]$ to the noise parameters $r_{\rm acc}$ and $r_{\rm oms}$ defined in Eq.~\eqref{eq:noise_amplitude_parametrization}. All prior distributions are summarized in Table~\ref{tab:prior_table}.

\begin{table}[!b]
    \centering
    \begin{tabular}{c c}
    \hline
     Parameter & Prior \\
     \hline
     $f_{\rm GW}$  & $\mathcal{U}(f_{\rm min}, f_{\rm max})$ \\
     $\phi_0$ & $\mathcal{U}(0, 2\pi)$ \\
     $\log_{10} A_0$ & $\mathcal{U}(-24, -20)$ \\
     $\sin \beta$ & $\mathcal{U}(-1, 1)$ \\
     $\lambda$ &  $\mathcal{U}(0, 2\pi)$ \\ 
     $\cos \iota$ & $\mathcal{U}(-1, 1)$ \\
     $\psi$ & $\mathcal{U}(0, \pi)$ \\
     $\dot{f}_{\rm GW}$ & $\mathcal{U}(\dot{f}_{\rm min}, \dot{f}_{\rm max})$ \\
     \hline
     $r_{\rm acc}$ &  $\mathcal{U}(10^{-1}, 10)$ \\
     $r_{\rm oms}$ &  $\mathcal{U}(10^{-1}, 10)$ \\
     \hline 
     \hline
    \end{tabular}
    \caption{Prior choice on the DWD parameters.}
    \label{tab:prior_table}
\end{table}

\subsection{Prior on the number of sources}
\label{sec:number_prior}

In trans-dimensional inference, the prior $\pi(\Theta|H)$, introduced in Eq.~\eqref{eq:BayesTheorem}, includes a contribution that depends explicitly on the number of sources in each population. It is reasonable to assume different species and individuals for each population  to be independent; hence we can factorize the prior as
\begin{equation}
    \pi(\Theta|H) = \prod_{\alpha=1}^{N_{\rm sp}} \left[\pi(N_{\rm pop,\alpha}|H) \prod_{i=1}^{N_{\rm pop,\alpha}}\pi(\theta_{i,\alpha}|H)\right] \, ,
\end{equation}
where the individual-source parameter priors are specified in Section~\ref{sec:Priors}. This is the prior we choose in our GF pipeline and the one adopted in all the test cases in Sections \ref{sec:SingleBin}, \ref{sec:Results}, \ref{sec:Stress_test}.

Our aim is to construct DWD catalogs suitable for subsequent population studies while limiting their dependence on a specific astrophysical population model. We therefore adopt a phenomenological source-count prior that penalizes additional sources, with a penalty motivated by their signal-to-noise ratio. For a species $\alpha$ with $N_{\rm pop,\alpha}$ individuals, we use the number prior
\begin{equation}
    \pi(N_{\rm pop,\alpha}|H) = \left(1-e^{-\rho_{\rm thr}^2/2}\right)\, e^{-N_{\rm pop,\alpha} \rho_{\rm thr}^2/2} \, . 
    \label{eq:general_aic_prior}
\end{equation}
For $\rho_{\rm thr}>0$, this is a normalized geometric distribution. Each additional source feels a log-prior penalty of $\rho_{\rm thr}^2/2$. As $\rho_{\rm thr}\rightarrow0$, the relative prior weights of any two finite source counts approach unity. For a fixed total birth rate, $R_{b,\alpha}$, and a birth proposal equal to the individual-source parameter prior, the death rate of the $j$-th individual of species $\alpha$ is
\begin{equation}
    R_{j,d,\alpha} = \frac{R_{b,\alpha}}{N_{\rm pop,\alpha}}e^{\rho_{\rm thr}^2/2+\delta\log\mathcal{L}_{j,d,\alpha}} \, , 
\end{equation}
where $\delta\log\mathcal{L}_{j,d,\alpha}$ is the variation of the likelihood due to the death of an individual. This follows from Eq.~\eqref{eq:birth_death_rates}, assuming a fixed birth rate and a birth proposal that coincides with the prior.

The choice of $\rho_{\rm thr}$ can affect both the inferred source count and the efficiency of posterior sampling. As discussed in Appendix~\ref{app:convergence and mixing}, when the parameter space is complicated, noise fluctuations and overlapping signals can induce spurious signals that cannot be associated with individual injected sources. These artifacts may change the source-count posterior or simply extend the burn-in phase. In Section~\ref{sec:SingleBin} and~\ref{sec:Stress_test} we show test cases for $\rho_{\rm thr}=3$, while in Section~\ref{sec:Results}, we present a GF inference for a uniform prior. We notice that the analysis becomes more complicated when the injections contain signals with SNR comparable with $\rho_{\rm thr}$.  A joint analysis of individually resolved sources and the unresolved foreground therefore requires careful treatment of the source-count prior and the allocation of signal power between these components.

\section{Efficient parallel computations for the Global Fit}
\label{sec:Parallel computations}

\subsection{Thread-safe shared-memory parallelization}
\label{sec:multithread}

As discussed in Section~\ref{sec:CTMCMC}, \texttt{samsara} computes birth, death, and mutation rates at each step of the CT Markov chain. These computations require sampling new points in the parameter space, evaluating new waveforms for deterministic signals or PSDs for stochastic signals, computing the posterior at the proposed points, and combining all these contributions in the calculation of the rates, Eq.~\eqref{eq:birth_death_rates}. Since these operations are independent across all processes and species, \texttt{samsara} is an intrinsically parallel algorithm, whose convergence can be accelerated by increasing the number of proposed points. In fact, \texttt{samsara} chooses the next state of the chain proportionally to the rates, and thus to the log-posterior; therefore, a larger number of proposed points means a faster exploration of the parameter space, provided the computational resources that can handle all the proposed points in parallel. We emphasize that the computations are mutually independent, as they read the shared CT-MCMC state without modifying it. This implies that \texttt{samsara} admits a straightforward multithread parallelization scheme, free of race conditions across all species and moves. 

To fully exploit parallelism, we avoid a multiprocess implementation, which would slow down the computations by copying the residuals and the covariance matrix at each likelihood evaluation. In a realistic GF for LISA, this overhead becomes prohibitive, because the number of parallel computations is typically $10^3$–$10^4$. 

For an efficient multithreaded implementation, we prefer the compiled languages C++/CUDA over Python, whose multithreading is prevented by the GIL for CPU-bound workloads such as ours. The main advantage of this choice is the finer and more flexible control over memory, which is important given the size of the parameter space in the GF and the trans-dimensional nature of the problem. The indexing scheme to slice the arrays for the computation of the likelihood described in Section~\ref{sec:likelihood} is fully exploited in C++. Furthermore, as we will show in the Section~\ref{sec:Amortized data structures for trans-dimensional samplers}, in \texttt{samsara++} we introduce new data structures, designed to avoid unnecessary allocations and de-allocations of memory when births and deaths take place.  

Therefore, the rates are computed in parallel in C++ with OpenMP and with batched CUDA kernels on the GPU. To saturate the available CPUs or GPUs, \texttt{samsara++} increases or reduces the number of proposed points, finding the optimal trade-off between collecting a large number of samples and achieving good mixing.

\subsection{Amortized data structures for trans-dimensional samplers}
\label{sec:Amortized data structures for trans-dimensional samplers} 

Given an efficient parallelization, the next task is to avoid reallocating memory every time the dimension of the parameter space changes. In \texttt{samsara++}, we address this by using indexed dynamic arrays. For each individual in the society, we store in a structure \texttt{state} the parameters, the log-prior values, and waveforms or PSDs. These data structures are implemented as flattened one-dimensional C++ vectors, which are preferred over nested vectors per species and individuals, due to their lower overhead and improved memory locality, resulting in better performance. Let us define this flattened array corresponding to a quantity $x$ as \texttt{x}. For instance, $\Theta$ denotes the vector of parameters in the current society, while \texttt{Theta} is its corresponding data structure in \texttt{samsara++}. For each species $\alpha$, we define its capacity as the maximum number of individuals whose memory is allocated in \texttt{state}, and we initialize it using
\begin{equation}
    N_{\max, \alpha} = {\rm max}\left(2, N_{\rm pop,\alpha}^{\rm ini}\right)\, , 
\end{equation}
where $N_{\rm pop,\alpha}^{\rm ini}$ is the initial number of individuals for the species $\alpha$. If $x$ has length $\ell_{x,\alpha}$ for an individual of species $\alpha$, in \texttt{state} we allocate storage in the flattened one-dimensional array \texttt{x} for \texttt{len$\_$x} entries computed according to
\begin{equation}
    \texttt{len\_x} = \sum_{\alpha=0}^{N_{\rm sp}-1}N_{\max,\alpha}\ell_{x,\alpha} \, , 
\end{equation}
with $N_{\rm sp}$ the number of species in the society. In total, we reduce the memory allocation from $\mathcal{O}(N_{\rm gen})$ to $\mathcal{O}\left(\log_2 N_{\rm pop,\alpha}^{\max}\right)$, where $N_{\rm gen}$ is the total number of CT-MCMC iterations and $N_{\rm pop}^{\max}$ is the largest  population size explored by the chain. This logarithmic resizing cost is the standard behavior expected from amortized dynamic-array resizing~\cite{Tajran}.

Once unnecessary memory allocations are avoided, an efficient indexing scheme to extract the parameters from the flattened arrays is needed. In fact, \texttt{x} contains not only the $x$-arrays of the individual in the current society, but also of individuals removed in previous steps, because $N_{\max,\alpha}\geq N_{\rm pop,\alpha}$. Therefore, if deaths occur, $x$-arrays of removed individuals could be in the middle of \texttt{x}. The choice of moving such components to the bottom of \texttt{x} for every species is suboptimal, because the complexity would be $\mathcal{O}\left(\ell_{x,\alpha}\right)$ at each accepted-death step. For this reason, we define \texttt{ind$\_$start$\_$x} as an array of length
\begin{equation}
    \texttt{len\_start} = \sum_{\alpha=0}^{N_{\rm sp}-1}N_{\max,\alpha}\, . 
\end{equation}
Such an array contains the starting index for the quantity $x$ in the flattened array \texttt{x}. For instance, the $k$-th parameter of the $j$-th individual of the species $\alpha$ is extracted according to
\begin{equation}
    \texttt{x}\left[\, k+\texttt{ind$\_$start$\_$x}\left[j+N_{\max,\rm cumul,\alpha}\right]\,\right]=x_{k,j,\alpha} \, ,
\end{equation}
where $N_{\max,\rm cumul,\alpha}$ is the cumulative sum of the capacity up to the $\alpha$-th species. In this way, the index $j$ runs from $0$ to $N_{\rm pop,\alpha}-1$ and iterates over the living individuals in the population, and \texttt{ind$\_$start$\_$x} skips the holes in \texttt{x} due to removed individuals, which have not been removed from the arrays to optimize the computational cost. \texttt{ind$\_$start$\_$x} is initialized by filling, for each species $\alpha$, the first $N_{\rm pop,\alpha}^{\rm ini}$ entries starting at offset $N_{\max,\rm cumul,\alpha}$ with the starting indices of the individuals' $x$-blocks in \texttt{x}, spaced by $\ell_{x,\alpha}$. Then, if the accepted move is a death of the $j$-th individual of the species $\alpha$, \texttt{ind$\_$start$\_$x} is simply updated by exchanging its element at $j+N_{\max,\rm cumul,\alpha}$ with the one at $N_{\rm pop,\alpha}-1+N_{\max,\rm cumul,\alpha}$ (the last one in the active population). Therefore, the complexity in dealing with death moves is $\mathcal{O}(1)$ in \texttt{samsara++}.

\begin{figure}[t!]
    \centering
    \includegraphics[width=1.0\columnwidth]{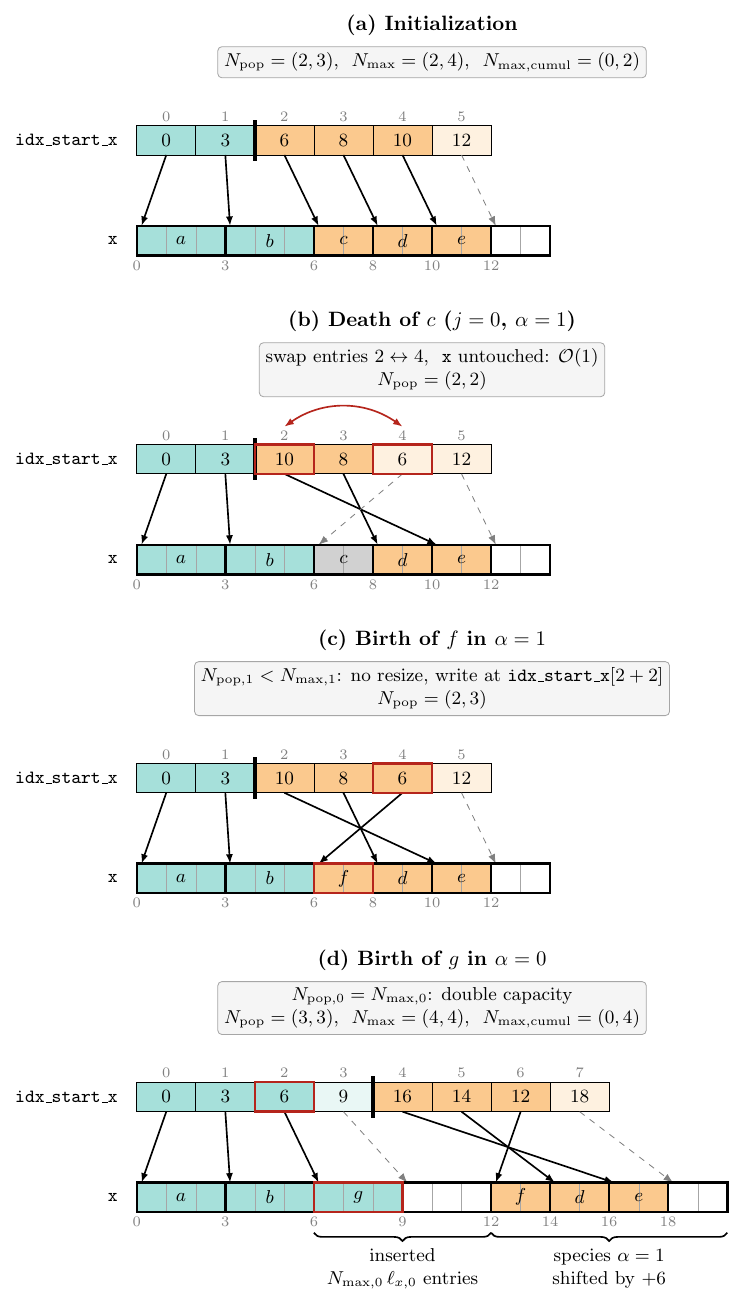}
    \caption{Indexing scheme of \texttt{samsara++} for two species characterized by a quantity $x$. The values of $x$ are stored in a single array with one block per species, which doubles in size whenever it fills up, so that the number of memory allocations grows logarithmically with the population size rather than linearly with the number of births. Alive individuals are then accessed through an index array; therefore a death only requires swapping indices, without moving the stored data.}
    \label{fig:indexing_scheme}
\end{figure}

In Figure~\ref{fig:indexing_scheme}, we summarize the initialization and update of the data structures used in \texttt{samsara++}.

\subsection{Parallel evolution steps for non-overlapping species}
\label{sec:parallel_species_evolution}

\texttt{samsara} evolves a society composed of multiple populations, each associated with a different species, using birth, death, and mutation processes. In the CT framework, this can be done with two prescriptions, which target the same posterior distribution but differ in how the chain is advanced and samples recorded. The first, adopted in~\cite{Astorino:2025ccl}, selects, via the Poisson dynamics described in Section~\ref{sec:CTMCMC}, a new state at each step of the chain. Each sample is associated with its expected waiting time, which is defined as the inverse of the total rate with respect to such a clock and represents the average lifetime of the state. In this scheme, the distribution of the samples weighted by the waiting times converges to the posterior distribution. In the second prescription, the samples are collected on a regular time grid, defined by a global clock $t_{\rm grid}$. Each step of the CT-MCMC now corresponds to a time interval of length $t_{\rm grid}$, within which the state evolves via Poisson dynamics. Unlike in the first prescription, the waiting time of each jump is drawn from the exponential distribution with rate equal to the total rate, and jumps are performed as long as the sum of the waiting times does not exceed $t_{\rm grid}$. In this approach, only the state at the end of each interval is added to the posterior samples, thus the samples carry no weights, and their distribution converges to the posterior.

The second scheme is particularly suitable for Gibbs-like sampling. Since evolving a species with the Poisson dynamics for a fixed time $t_{\rm grid}$ leaves the posterior invariant, individual species can be evolved in sequence, and a sample is collected after a full sweep over all the species. For the Gaussian likelihood described in Eq.~\eqref{eq:Whittle}, signals in disjoint frequency ranges have zero cross product, and the likelihood factorizes into separate contributions. Thus, if the prior is separable over species, populations whose signals do not overlap in frequency are conditionally independent given the rest of the society, and can be evolved in parallel within a Gibbs-like step. This is a great speed up in the exploration of the trans-dimensional parameter space and the collection of samples at fixed dimension.

In our GF pipeline, we divide the LISA frequency band into $N_{\rm bin}$ bins, considering DWDs in different bins as separate species. The spacing of the bins in \texttt{samsara++} can be uniform or logarithmic between $f_{\rm min}$ and $f_{\rm max}$, defined in Section~\ref{sec:likelihood}, or adaptive, where the size of the bins is specified by the user. A frequency bin $I$ is characterized by its central frequency $f_{\rm bin}^{(I)}$ and its width $\Delta f_{\rm bin}^{(I)}$. A sufficient condition for the independence between two bins $I$ and $J$, with $f_{\rm bin}^{(I)}<f_{\rm bin}^{(J)}$, is that the waveform emitted by the DWD with the highest frequency in the lower bin, $f_{\rm bin}^{(I)}+\Delta f_{\rm bin}^{(I)}/2$, does not overlap with the one generated by the DWD with the lowest frequency in the upper bin, $f_{\rm bin}^{(J)} - \Delta f_{\rm bin}^{(J)}/2$. For observations of duration $T_{\rm obs}$, we estimate the maximum signal bandwidth in the interval \(I\) as 
\begin{equation}
    \begin{split}
        \Delta f_{\rm GW,\max}^{(I)} =& \frac{2}{T_{\rm obs}}\left[4+2\pi \left(f_{\rm bin}^{(I)}+\frac{\Delta f_{\rm bin}^{(I)}}{2}\right)\frac{R_{\rm ES}}{c}\right]\\
        & + \dot{f}_{\rm GW,\max} T_{\rm obs}\, ,
    \end{split}
\end{equation}
where $\dot{f}_{\rm GW,\max}$ is the maximum frequency derivative allowed by the prior, $R_{\rm ES}$ is the Earth–Sun distance. The condition for independence between the bins can be summarized as
\begin{equation}
    f_{\rm bin}^{(I)}+\frac{\Delta f_{\rm bin}^{(I)}}{2} +5\, \Delta f_{\rm GW,\max}^{(I)}\leq f_{\rm bin}^{(J)} - \frac{\Delta f_{\rm bin}^{(J)}}{2} - 5\, \Delta f_{\rm GW,\max}^{(J)}\, . 
\end{equation}
The factor of five provides an additional margin around the estimated signal bandwidths. Although this is only a sufficient condition, given the large number of bins used in our analysis, it is enough to fully exploit the CPUs or GPUs in the parallel computation across the bins. Details on the bin choice used for our analysis will be given in Section~\ref{sec:Results}.

If we define $N_{\rm CPU}$ as the number of available CPUs and $N_{\rm batch, GPU}$ the number of batched calculations done on the GPU, a single step of the parallel-Gibbs \texttt{samsara} is equivalent to $N_{\rm CPU}$ or $N_{\rm batch, GPU}$ steps of \texttt{samsara} where DWD evolves. The parallel-Gibbs mutation steps increase by $N_{\rm CPU}$ or $N_{\rm batch, GPU}$ the mixing between the chains at fixed dimensions, allowing to sample in a reasonable amount of time the parameters of $10^4$ DWDs.

\subsection{Parallel tempering}
\label{sec:Parallel tempering} 

During the burn-in, for trans-dimensional samplers, it is natural to go through an initial phase where the number of sources exceeds the number of injected signals; examples are presented in~\cite{Astorino:2025ccl}. This behavior can become more pronounced and problematic for high-SNR signals, since it is more difficult to enter the main mode of the posterior. In such cases, \texttt{samsara} may fit loud signals with an excess of spurious sources of similar frequencies and smaller amplitudes, if the proposal is not optimal. The spurious signals are removed over several iterations of the chains, but, for high SNR signals, these steps would require too much time for a full GF analysis.

To mitigate this behavior, like other GF pipelines~\cite{Littenberg:2023xpl,Katz:2024oqg,Deng:2025wgk}, \texttt{samsara++} employs PT~\cite{Wang_2005,Earl_2005}. PT evolves an ensemble of chains at different temperatures. At higher temperatures, the flattened likelihood landscape facilitates transitions between posterior modes. Exchanges between temperatures allow the cold chain, which targets the desired posterior, to escape configurations containing redundant sources and explore more probable configurations, reducing the prolonged burn-in associated with high-SNR signals.

Our sampler supports two complementary tempering schemes, which we refer to as internal and external tempering. Internal tempering uses Neal’s tempered-transition method~\cite{Neal1996SamplingFM} within a mutation move. A candidate state is generated by traversing a temperature ladder from the current temperature to higher temperatures and back. The complete transition is then accepted or rejected using a path-dependent acceptance probability that preserves detailed balance at the starting temperature. In external PT, several CT Markov chains run with different temperatures, and swaps between chains are proposed every fixed number of steps. This corresponds to the usual PT, with the novelty that the parameter space of the chains can differ and the implementation is done for CT-MCMC. Within \texttt{samsara++}, the temperature ladder can be uniformly or geometrically spaced between $T_{\rm min}$ and $T_{\rm max}$, or adapted during the burn-in. For the adaptive ladder we follow the implementation of~\cite{Vousden_2015}, used also by \texttt{Eryn}~\cite{Karnesis:2023ras}, where the temperature dynamics is defined in terms of the quantity
\begin{equation}
    S_i = \log (T_i - T_{i-1}) \, .
\end{equation}
In order to achieve a uniform acceptance ratio of swaps between different chains, $S_i$ is driven in the burn-in phase by the hyperbolic decay
\begin{equation}
    \frac{\dd S_i}{\dd t} = k(t) [A_i(t) - A_{i+1}(t)] \,, \quad k(t) = \frac{1}{\nu}\frac{t_0}{t + t_0} \, ,
\end{equation}
where $A_i$ is the acceptance for the swap of the chain at temperature $T_i$ and $k$ is the instantaneous decay constant for ladder adjustments. The initial amplitude of adjustment $\nu$ and the time-scale of temperature adjustment $t_0$ are chosen by the user according to the prescriptions given in~\cite{Vousden_2015}. 

The importance of PT is discussed further in Appendix \ref{app:convergence and mixing}.

\subsection{Postprocess}
\label{sec:Postprocess}

\texttt{samsara++} inherits the diagnostic and post-process infrastructure implemented in \texttt{samsara}~\cite{Astorino:2025ccl}, which includes optimized sample-storage scheme, avoiding writing the full society to the output when only a few individuals change. The samples are saved in a HDF5 file, which is read and unpacked directly in \texttt{C++} to reconstruct the sampled societies. For the GF analysis presented here, this storage scheme limits the output file size to a maximum of \(20\,\mathrm{GB}\).

Once the samples are read from the HDF5 file, the identification of signals, together with their associated detection probabilities, is handled by the \texttt{PETRA} algorithm~\cite{Johnson:2025oyu}. Once a catalog is generated from the posterior samples, we find the best match of the injections with the single-source posteriors in the catalog by applying the Hungarian algorithm\footnote{This is equivalent to a single step of \texttt{PETRA} where the catalog mock distributions are fixed and cannot vary.}. To make this procedure compatible with the \texttt{samsara++} indexing scheme and sample storage, we have entirely rewritten \texttt{PETRA} in C++, improving the code's efficiency through alternative data structures, including a straightforward OpenMP parallelization. This choice is particularly helpful, because, for a GF run, the number of samples and size of the parameter space are huge; without the efficient memory handling, speed, and parallelization of C++, the post-processing would be challenging in Python\footnote{In \texttt{samsara++}, all the credits of the algorithm to extract the catalogs from the GF samples are given to \texttt{PETRA} and the authors of~\cite{Johnson:2025oyu}.}.

\begin{figure*}
    \centering
    \includegraphics[width=0.75\linewidth]{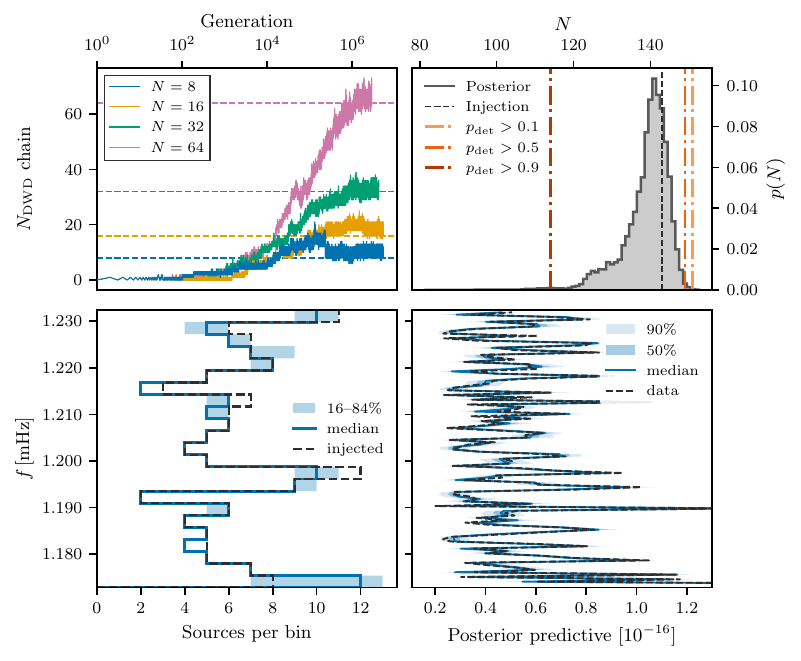}
    \caption{Inference on the single frequency bin $[1.1728, 1.2324]\,\rm mHz$. Top left: trace of the number of DWDs as a function of the generation, for PT runs with 8, 16, 32, and 64 injected sources (dashed horizontal lines), starting from an empty state. Top right: posterior on the number of DWDs (grey) for 143 injections (dark grey dashed line), and number of sources in the \texttt{PETRA} catalog with $p_{\rm det}$ greater than 0.1, 0.5, and 0.9 (orange dot-dashed lines, from light to dark). Bottom left: number of sources in 23 equally spaced frequency sub-bins; the blue line and the light blue band are the median and the 16--84\% credible interval over the societies explored by \texttt{samsara++}, while the dark grey dashed line is the injected population. Bottom right: posterior predictive of the modulus of the $A$-channel data, $|\tilde d_A(f)|$, over the same frequency range as the bottom-left panel. The blue line is the median, the dark and light shaded regions mark the 50\% and 90\% credible intervals, and the dark grey dashed line is the data. Both the posterior predictive draws and the data are shown as the root mean square of $|\tilde d_A|$ over 16 adjacent frequency bins ($\simeq 0.5\,\mu\mathrm{Hz}$).}
    \label{fig:singlebin_comparisons}
\end{figure*}

\section{Test case of DWDs in a single bin}
\label{sec:SingleBin}

Before applying \texttt{samsara++} to the full DWD catalog, we perform simpler test cases to check the accuracy of \texttt{samsara++} and its scalability with the number of sources. We also evaluate the effectiveness of the trans-dimensional PT scheme described in Section~\ref{sec:Parallel tempering}. Here we focus on the trans-dimensional inference, while in Appendix~\ref{app:ppplot}, we validate the sampler at fixed dimensionality.

We consider 143 DWDs with SNR greater than $5$ in the frequency bin $[1.1728, 1.2324]\, \rm mHz$, adopting the parameter priors described in Section~\ref{sec:Priors}. To ensure an efficient exploration of the parameter space at fixed dimensions, the inference is performed by using a birth rate fixed to unity and a mutation rate equal to the number of individuals in the population. As discussed in Appendix~\ref{app:convergence and mixing}, PT is a crucial ingredient to obtain convergence and good mixing. We therefore use it in all runs, choosing $T_{\rm min}=1$ and $T_{\rm max}=15$ for the scalability in Section~\ref{sec:Scalability with the number of injections} and $T_{\rm max}=30$ for the full-bin analysis in~\ref{sec:Inference on the full bin}. We adopt the number prior defined in Eq.~\eqref{eq:general_aic_prior} with $\rho_{\rm thr}=3$, to reduce the presence of spurious sources in the sampler due to the complicated structure of the parameter space in the presence of overlapping sources. Appendix~\ref{app:convergence and mixing} examines the influence of this choice on the inferred source count and sampling efficiency, showing that a uniform source-count prior leads to less efficient sampling for the injected population considered here.

\subsection{Scalability with the number of injections}
\label{sec:Scalability with the number of injections}

In order to test the performances of the code as a function of the number of sources, we select subsets containing $\{8, 16, 32, 64\}$ DWDs in the bin and we perform a Bayesian inference on each sub-group. Every run starts from an empty society, without using the injected source parameters to initialize the sampler.

In the upper left panel of Figure~\ref{fig:singlebin_comparisons}, we plot the (source-count) traces obtained using PT with 8 temperatures. In all four runs, the inferred count approaches the injected values after approximately  $10^6$ iterations, and then fluctuates around a slightly larger value. As shown in~\cite{Astorino:2025ccl}, the small fluctuations in the number are due to births and deaths of spurious sources fitting the noise realization and that do not have a large detection probability when a catalog is constructed from the posterior samples. An interesting result is that \texttt{samsara++} recovers different numbers of injections in approximately the same number of generations. This happens because when the number of sources is larger, the sampler needs more iterations to get all of them, but at the same time it is easier to propose births of valid points, since the parameter space is more populated. 

\subsection{Inference on the full bin}
\label{sec:Inference on the full bin}

We perform an inference over all the 143 DWDs in the full frequency bin, using PT with 16 temperatures. The injected SNRs range from \(5.8\) to \(62.9\), with a median of \(11\). Among the 143 injected sources, 139 have SNRs larger than 8, while 4 have SNR between 5 and 8. The code performs approximately 125 iterations$/s$ per chain with 64 CPUs, and, since the computations over the 16 chains are parallelized, the effective rate of iterations with \texttt{samsara++} is 2000 iterations$/s$. The burn-in consists of approximately $5\times 10^6$ iterations performed in 6 hours. We then collect the samples  over a further $10^6$ generations, requiring approximately one additional hour.

\begin{figure*}
    \centering
    \includegraphics[width=\linewidth]{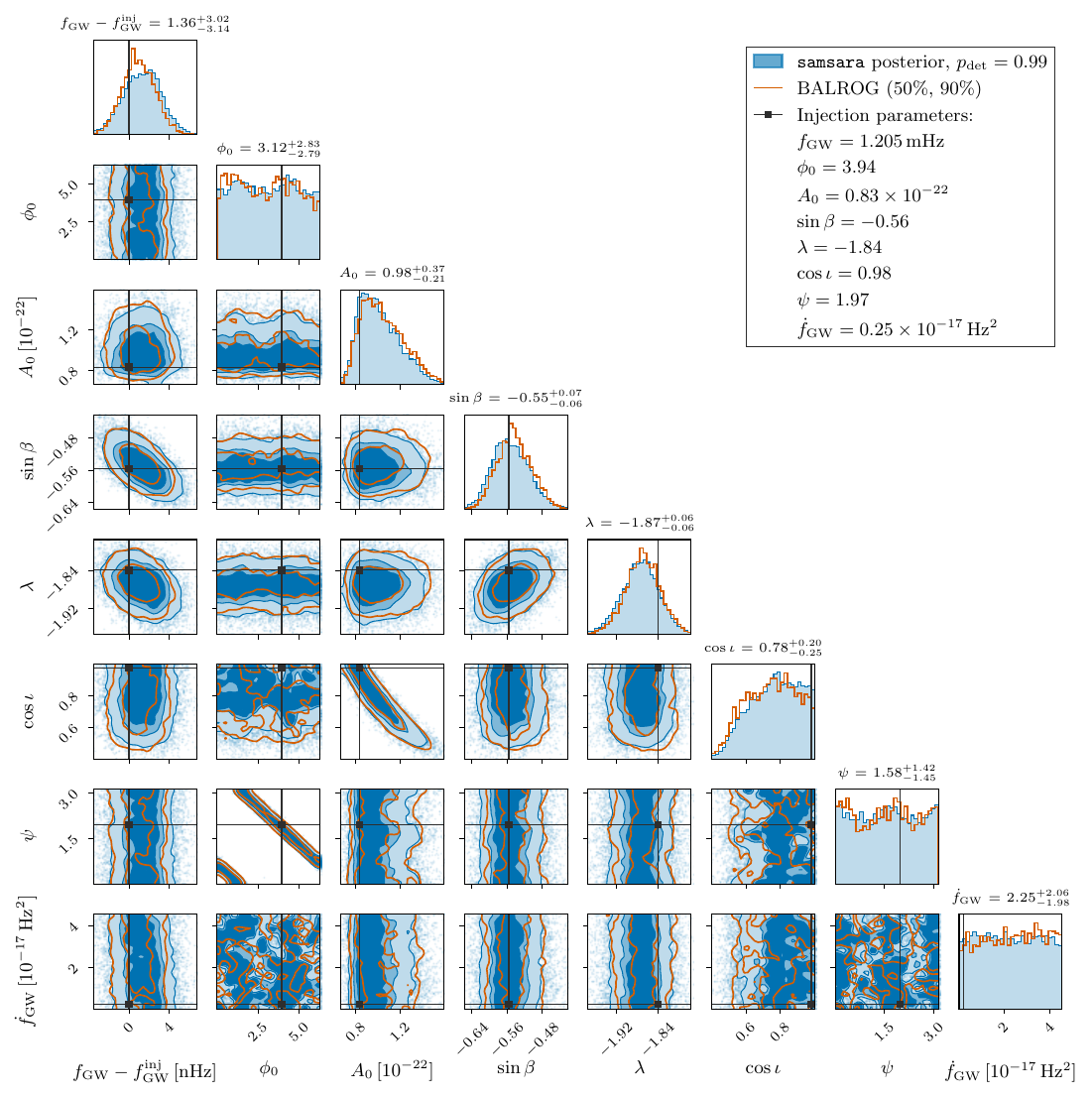}
    \caption{
    Marginal posterior distributions of the parameters of a DWD in the \texttt{PETRA} catalog extracted from the \texttt{samsara++} posterior samples of the test case presented in Section~\ref{sec:SingleBin}. The source has detection probability $p_{\rm det}=0.99$ and is matched to an injected DWD, whose parameters are marked by the black lines and listed in the legend. In the 2D marginals, the shaded regions mark the 50\%, 68\%, and 90\% credible regions. For comparison, the orange lines show the posterior of an independent single-source analysis with \texttt{BALROG}~\cite{2019PhRvD.100h4041B,Roebber:2020hso,Finch:2022prg,Moore:2023nbw}, in which the injected DWD alone is added to the same noise realization of the \texttt{samsara++} run, and the likelihood includes instrumental noise only (no other DWDs and no confusion noise). This run uses the same observation time and frequency grid, uniform priors over the plotted ranges, and nested sampling with \texttt{dynesty}~\cite{Speagle:2019ivv}. Orange contours mark the 50\% and 90\% credible regions.
    }
    \label{fig:singlebin_countours}
\end{figure*}

In the upper-right panel of Figure~\ref{fig:singlebin_comparisons}, we plot the posterior distribution on the number of sources obtained with the inference and the number of sources in the catalog reconstructed with the C++ port of \texttt{PETRA} with $p_{\rm det}$ greater than 0.1, 0.5, and 0.9. Remarkably, the posterior on the number is peaked around 138, corresponding almost to the number of injected sources with SNR greater than 8. The left tail of the posterior corresponds to injected sources with large SNR and small death rate that can be removed and added on the society with tiny probability. The right tail of the distribution includes spurious sources used to fit the noise and overlapping sources, because of the complicated structure of the $\mathcal{O}(10^3)$-dimensional parameter space. As expected, the catalog sources with $p_{\rm det}>0.1$ exceed the number of injections, consistently with the inclusion of low-probability components associated with noise fitting and source confusion. Sources selected with $p_{\rm det} \geq 0.9$ show stronger posterior samples and agree with the injected signals. As an example, we plot in Figure \ref{fig:singlebin_countours} the contours of the second DWD in the \texttt{PETRA} catalog, with $p_{\rm det}\geq 0.99$. The plot shows a very good quality of the posterior samples obtained with \texttt{samsara++}, due to PT and efficient proposal cycles used for DWD mutations, described in Appendix~\ref{app:proposals}. In addition, we compare the \texttt{samsara++} contours with the ones obtained with a single-source run using \texttt{BALROG}~\cite{2019PhRvD.100h4041B,Roebber:2020hso,Finch:2022prg,Moore:2023nbw} and nested sampling as implemented in \texttt{dynesty}~\cite{Speagle:2019ivv}, finding excellent agreement between the results and accuracy in reconstructing the injected values. In the bottom-left panel of Figure~\ref{fig:singlebin_comparisons}, we also compare the posterior credible intervals for the number of sources in frequency sub-intervals with the corresponding injected counts. In this case, the sampler recovers the frequency distribution of DWDs, up to few contributions due to the 4 sources with SNR smaller than 8. In the bottom right panel of Figure~\ref{fig:singlebin_comparisons}, we plot the posterior predictive of the modulus of the A-channel signal, obtained as the root mean square of $|\tilde{d}_A|$ over 16 adjacent frequency bins. The comparison between the 50\% and 90\% percentiles of the posterior predictive with the data shows an excellent agreement between the \texttt{samsara++} posterior samples and the injection.

\section{Mock GF of DWDs and instrumental noise}
\label{sec:Results}

\subsection{Injection}
\label{sec:Injection}

The current search pipelines for DWDs, see, e.g.,~\cite{Cornish_2005, Crowder_2007, Prix_2007, Strub:2024kbe}, are expected to reconstruct the parameters that maximize the likelihood of the loudest sources in the catalog. These estimates could be used to initialize \texttt{samsara++} near well-supported source configurations, reducing the prolonged burn-in associated with representing high-SNR signals through redundant components\footnote{Such initialization is done also in other pipelines~\cite{Katz:2024oqg}.}. To assess performance on a larger population without the very loud signals responsible for this difficulty, here we consider a test case containing 866 sources with $5\leq \rm SNR \leq 30$ in the $\rm mHz$ band starting from 0 DWDs in the inference. This test assesses the potential of \texttt{samsara++} for joint trans-dimensional inference of DWDs and instrumental noise. 

The DWD parameters are extracted from the catalog described in Section~\ref{sec:Astrophysical Model} selecting sources potentially resolvable by LISA after one year of observations. We then inject Gaussian instrumental noise using the model described in Section~\ref{sec:Instrumental noise}. To assess the sampler’s efficiency under controlled conditions, this preliminary analysis omits the unresolved DWD foreground, temporal variations in the instrumental noise PSD, and signals from MBHBs and EMRIs.

We divide DWDs in 1000 frequency bins, following the prescription described in Section~\ref{sec:parallel_species_evolution}, and adopt the source-parameter priors defined in Section~\ref{sec:Priors}. For the prior on the number, here we use a uniform prior to be completely agnostic on the number of sources. The proposal distributions used in this analysis are illustrated in Figure~\ref{fig:proposals}. Finally, we set the algorithm with adaptive birth and death rates.

\begin{figure*}
    \centering
    \includegraphics[width=\linewidth]{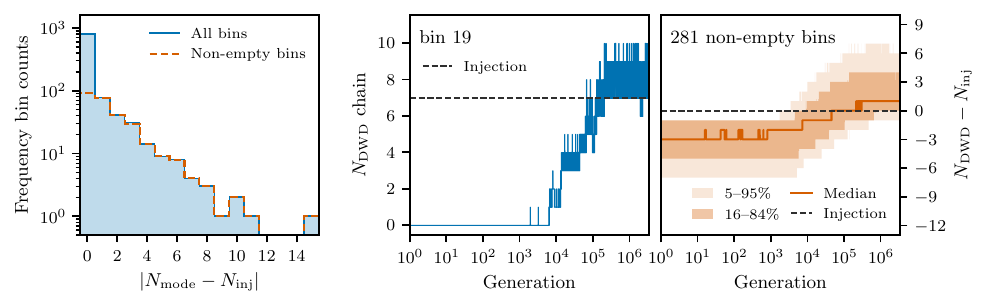}
    \caption{Left panel: histogram of the absolute difference between the number of injected sources and the posterior mode of the number of sources in each frequency bin, for all the 1000 bins (blue) and for the 281 bins containing at least one injected source (orange, dashed). Middle panel: trace of the number of sources as a function of the generation for the 19th frequency bin; the dashed line marks the injected number. Right panel: median and 16--84\% and 5--95\% bands, across the 281 non-empty bins, of the difference between the number of sources and the injected one as a function of the generation.}
    \label{fig:number_DWD}
\end{figure*}

\subsection{Convergence to the true number of injections}
\label{sec:number_convergence}

We start by assessing how well \texttt{samsara++} recovers the number of injections. In the left panel of Figure~\ref{fig:number_DWD} we plot the absolute difference between the number of injected signals in each frequency bin and the posterior mode of the corresponding source count. These modes are computed after removing the burn-in phase and thinning the samples, as discussed in ~\cite{Astorino:2025ccl}. The close agreement between the inferred and injected counts demonstrates the ability of \texttt{samsara++} to infer the population size across the analyzed frequency range. This is a central result of this work: our algorithm starts from an \textit{empty initial state} and does not rely on any astrophysically informed priors, as shown in Sections~\ref{sec:Priors},~\ref{sec:number_prior}. In the middle panel of Figure~\ref{fig:number_DWD}, we plot the source-count trace of the DWDs in the 19th frequency bin. We note that \texttt{samsara++} keeps adding and removing individuals until the population stabilizes at the true number of sources. After that, the small fluctuations in the number correspond to fits of the noise realization with signals that have low detection probability in the catalog. In the right panel of Figure~\ref{fig:number_DWD}, we plot the percentiles (with respect to the bins) of the difference between the number of DWDs present at a given step and the injection. The median of such difference begins with a negative number, because here \texttt{samsara++} is initialized with an empty state, and terminates with a positive number, since some bins have DWDs that are still overfitting high-SNR sources and contain fits of noise fluctuations.

\subsection{Reconstruction of the injected total signal}

In Section~\ref{sec:number_convergence}, we assessed the capability of \texttt{samsara++} to estimate the number of injected signals. This inference, together with providing a catalog, depends on the choice of the priors. Moreover, source counts and parameters are correlated: configurations with different numbers of sources can produce similar fits to the data. For this reason, we introduce a metric to validate our inference that is not sensitive to such a degeneracy. For the Gaussian likelihood defined in Eq.~\eqref{eq:Whittle}, a natural choice is the reconstruction of the total deterministic signal, defined as
\begin{equation}
        h_{I,\rm tot}(\Theta_d, f) \equiv \sum_{\alpha=1}^{N_{\rm sp}^{\rm d}}\sum_{i=1}^{N_{\rm pop,\alpha}^{\rm inj}}h_I\left(\theta_{i,\alpha},f\right)\, .
    \label{eq:total_signal}
\end{equation}
Comparing the reconstructed total signal with its injected counterpart assesses how accurately the model recovers the combined deterministic contribution to the data. Furthermore, $h_{I,\rm tot}$ maps any society onto a common signal space, allowing a consistent comparison across models of different dimensions.

\begin{figure*}
    \includegraphics[width=\linewidth]{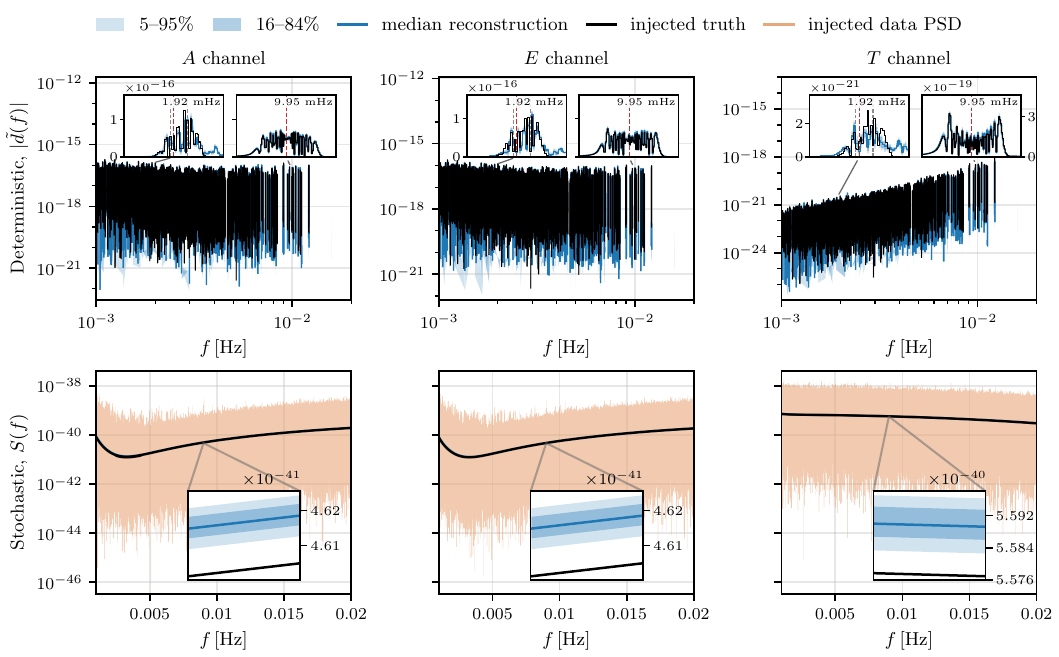}
\caption{Norm of the deterministic signal Fourier transform, $|\tilde{d}(f)|$ (top row), and PSD (bottom row) for the $A$ (left), $E$ (middle), and $T$ (right) channels; the PSD panels share the same axes. The blue line is the median obtained with \texttt{samsara++}, while the black line represents the injected value. The dark and light shaded regions mark the 16--84\% and 5--95\% credible intervals, respectively, and the orange line is the PSD of the injected data. In the top row, the bands are the credible intervals of $|\tilde{d}|$ itself, computed frequency by frequency in each channel. The deterministic component is identically zero outside the frequency windows around each source; these bins are omitted on the logarithmic axis. The insets in the top row show $|\tilde{d}|$ on a linear scale, bin by bin, around the injected frequency $f_0$ (red dashed line) of two sources. The left inset shows a source at $f_0 = 1.916\,\mathrm{mHz}$ with SNR $8.4$, whose signal overlaps with those of two other injected sources, at $-0.06$ and $+0.25\,\mu\mathrm{Hz}$ from it (grey dashed lines), within $\pm 0.89\,\mu\mathrm{Hz}$ of $f_0$. The right inset shows an isolated source at $f_0 = 9.955\,\mathrm{mHz}$ with SNR $24.1$, within $\pm 1.7\,\mu\mathrm{Hz}$ of $f_0$, whose injected signal is recovered bin by bin within the credible intervals over its full $\sim 2\,\mu\mathrm{Hz}$ support. The insets in the bottom row zoom in on the PSD around $f \simeq 9\,\mathrm{mHz}$.}
    \label{fig:total_signal}
\end{figure*}

In the top panels of Figure~\ref{fig:total_signal}, we plot the modulus of the total reconstructed signal in the $A$, $E$, and $T$ channels. Under the assumptions adopted in Section~\ref{sec:likelihood}, the $T$ channel response is almost insensitive to GWs at low frequencies. The plots show excellent agreement between the \texttt{samsara++} results and the injected signals within the credible intervals across all frequency bins, confirming the robustness of our algorithm.

The bottom panels show the instrumental noise PSDs reconstructed from posterior samples of the noise parameters. The inferred PSDs exhibit a small upward offset relative to the injected model. One possible explanation is the incomplete reconstruction of the deterministic signal; it could be that a coherent power remained in the residuals can be partly absorbed into the inferred noise PSD. Correlations between source and  noise parameters, together with challenges  with the decomposition of overlapping signals, may also affect the reconstruction. An additional explanation could be our restriction to $f_{\rm min}>1 \, \rm mHz$, which could limit the frequency coverage available to constrain the noise components. Extending the analysis to lower frequencies could provide additional information, particularly on the noise, but, at the same time would require consistent modeling of the Galactic foreground. In a future work, we plan to investigate a joint source, noise and foreground inference over a broader frequency range and assess whether longer runs and improved tempering  could reduce the observed discrepancy.

\section{Stress test on extremely high dimensional parameter space}
\label{sec:Stress_test}

As a final benchmark, we test the performance of \texttt{samsara++} in a parameter space containing more than $10^5$ source parameters. We simulate one year of LISA data sampled at 24 s of cadence and we inject all the sources in the catalog above 1 mHz with SNR $\ge 6$, for a total of 16435 injected DWDs. With eight parameters per DWD, the source parameter space has 131,480 dimensions at the injected population size. We divide the LISA frequency band into 250 bins, assigning a species to each frequency bin, as described in Section \ref{sec:parallel_species_evolution}. We adopt the parameter priors listed in Table~\ref{tab:prior_table} and the source-count prior with $\rho_{\rm thr}=3$, as in Section \ref{sec:SingleBin}. We also fixed the birth rate to unity. Since we start the inference from the injection, we do not use PT and we assume that the burn-in phase has been already reached. This test case provides an idealized benchmark for a future implementation of a preliminary search in the pipeline that would initialize the sources close to their true values, therefore we can just focus on showing the potential of \texttt{samsara++}.

\begin{figure*}
    \centering
    \includegraphics[width=0.75\linewidth]{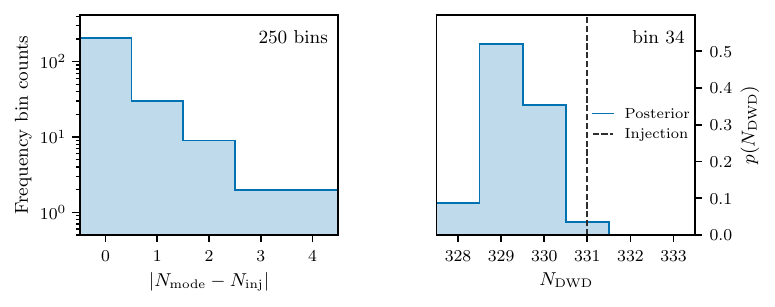}
    \caption{Left panel: histogram of the absolute difference between the number of injected sources and the posterior mode of the number of sources in each frequency bin. Right panel: posterior on the number (blue) and number of injections (dashed black) of the most populated frequency bin.}
    \label{fig:16_k_results}
\end{figure*}

We first note that all the frequency bins sample from a parameter space consistent with the injected values and SNRs. In particular, the left panel of Figure \ref{fig:16_k_results} shows that there are 2 bins with the mode of the number posterior shifted by 3 and 4, 10 bins with a shift of 2, around 30 bins with a shift of 1, and approximately 200 bins with the mode equal to the injection. In the right panel of Figure~\ref{fig:16_k_results}, we show the recovered number posterior for the most populated bin. The posterior is consistent with the injection of 331 DWDs.

In terms of efficiency, the code performs 15 million iterations in 16 hours on 8 AMD Ryzen 7 7735HS with Radeon Graphics cores (2 threads per core), corresponding to a sampling rate of 260.4 iterations$/s$. Furthermore, \texttt{samsara++} uses approximately 3 GB of RAM and produces 20 GB of stored samples.

\section{Discussion}

We have presented \texttt{samsara++}, an optimized C++/CUDA sampler for trans-dimensional inference built on a previous Python version, \texttt{samsara}, with new features and design choices specifically tailored to the LISA GF problem. \texttt{samsara++} is based on birth-death and mutation processes in the CT-MCMC framework, which represents a paradigm shift with respect to RJ-MCMC and product-space methods. The algorithm chooses whether to change dimension or to add, remove, or mutate individuals according to Poisson rates connected to posterior ratios between the current state and a set of proposed points with different dimensionality. Therefore, the Markov chain follows a posterior-driven trajectory in the trans-dimensional parameter space, resulting in an accurate and efficient sampling scheme. 

\texttt{samsara++} relies on custom data structures designed for GF inferences, in which the arrays storing parameters and signals change size during sampling, and many independent computations are performed in parallel. Thanks to lists of indices that keep track of living and dead individuals in the society, \texttt{samsara++} amortizes memory allocation by three orders of magnitude and reduces the complexity of removing a DWD from $\mathcal{O}(N_{\rm ch} N_{\rm het})\approx \mathcal{O}(10^3)$ to $\mathcal{O}(1)$. The parallelism is then organized in three nested levels: across chains, with a novel PT scheme adapted to CT-MCMC, across independent species evolving simultaneously within each chain, with a Gibbs sampling adapted to CT-MCMC, and across the computation of the Poisson rates of each evolving species.\footnote{The last case is the special case of the second with a single evolving species.} We parallelize with OpenMP on CPU and with batched CUDA kernels on GPU. Our C++/CUDA implementation enables thread-safe and memory-shared parallelization, where the residuals, covariance, and waveforms of the current state are not duplicated for the thousands of parallel computations of the posterior in the proposed points.  

We show the performance of \texttt{samsara++} on a first test case of 143 DWDs in a frequency bin of width $0.06\, \rm mHz$. Despite the high dimensionality of the parameter space and the large priors, the sampler recovers the injected number of sources for sub-groups of 8, 16, 32, and 64 sources in the bin in approximately $10^6$ iterations. The number of steps to converge does not depend on the number of injections, because large populations have more individuals to be found, but also a larger probability of proposing good ones. For the full inference of all the 143 DWDs in the bin, we observe a rate of 2000 iterations$/s$, with $10^6$ iterations needed by the sampler to reach the main mode of the posterior on the number. After the burn-in, we collect another $10^6$ samples, which reproduce a number posterior in agreement with the injection. The associated catalog built with \texttt{PETRA} contains good posterior samples on individual sources that match the injections. We remark that the results were obtained starting from zero signals and without assuming any information about the injection or the astrophysical population of DWDs. Moreover, to find DWDs, we only use PT and birth proposals, which are efficient enough to spot the signals in approximately one hour. Therefore, in this work, we do not rely on preliminary searches in the data. 

We then perform a more realistic inference of 866 DWDs with SNR between 5 and 30 and instrumental noise. We neglect high-SNR signals because their fit could be optimized with search pipelines discussed in the literature, while we focus on trans-dimensional analysis and parameter estimation. In this case, since we are not considering very loud DWDs, we do not make use of PT. We neglect sub-threshold sources, leaving a more detailed discussion about fitting unresolved signals with both birth-death processes and a parametric formula for the foreground for future work. DWDs are categorized into 1000 linearly-spaced frequency bins, which are evolved in parallel with \texttt{samsara++}. After $10^6$ iterations, 80\% of the bins match the number of the injected sources, while $10\%$ of the bins differ by one source, and the rest of the bins differ by two or more sources. The bins with an excess of signals show a decreasing trend in the number trace, meaning that the run has not fully converged yet.      

As a final benchmark, we test \texttt{samsara++} on a population of 16,435 DWDs drawn from the catalog described in Section~\ref{sec:Astrophysical Model}, corresponding to 131,480 source parameters at the injected population size. We initialize the sampler at the injected parameters, providing an idealized proxy for a future pipeline initialized using preliminary search results. The posterior mode of the source count matches the injected count in approximately $83\%$ of the 250 bins, with the largest discrepancies being three and four sources. The runs are achieved on 8 AMD Ryzen 7 7735HS with Radeon Graphics cores (2 threads per core) that performed 15 million iterations in approximately 16 hours, using 3 GB of RAM and producing 20 GB of stored samples.

In the near future, we pursue the direction of tackling the LISA Global Fit problem with \texttt{samsara++}, by including the Galactic binaries, the unresolved Galactic foreground, MBHBs, and EMRIs in the source model of our sampler. We will assess these developments using increasingly realistic simulated datasets, including the LDC-2a ``Sangria'' data challenge.

\begin{acknowledgments}

The authors are grateful to A. Klein for the multiple helpful conversations and for providing the waveforms used in this work. The authors thank S. Babak, E. Barausse, M. Crisostomi, G. Demasi, C. J. Moore, C. Pitte, J. Pomper, R. Rosso, A. Sesana, and the participants of the conference ‘‘Listening to the Cosmos: New Frontiers in Gravitational Wave Physics'' held at Galileo Galilei Institute for Theoretical Physics for useful discussions and comments. L.~V. acknowledges financial support from the project “LISA Global Fit’' funded by the ASI/Università di Trento Grant No. 2024-36-HH.0 - CUP F63C24000390001. RB acknowledges support from the ICSC National Research Center funded by NextGenerationEU, and the Italian Space Agency grant Phase B2/C activity for LISA mission, Agreement n.2024-NAZ-0102/PER. 

R. B. acknowledges computational resources provided by the European Union's H202 ERC Advanced Grant "PINGU". Additional computational work was performed using University of Birmingham BlueBEAR High Performance Computing facility and CINECA with allocations through INFN, Bicocca, and ISCRA project HP10BEQ9JB.

\end{acknowledgments}

\clearpage

\appendix

\input{appendix}

\clearpage

\bibliographystyle{apsrev4-2} 
\bibliography{main}

\end{document}

%% file: appendix.tex
\section{Proposals}
\label{app:proposals}

\begin{figure*}[p]
    \centering
    \includegraphics[width=\textwidth]{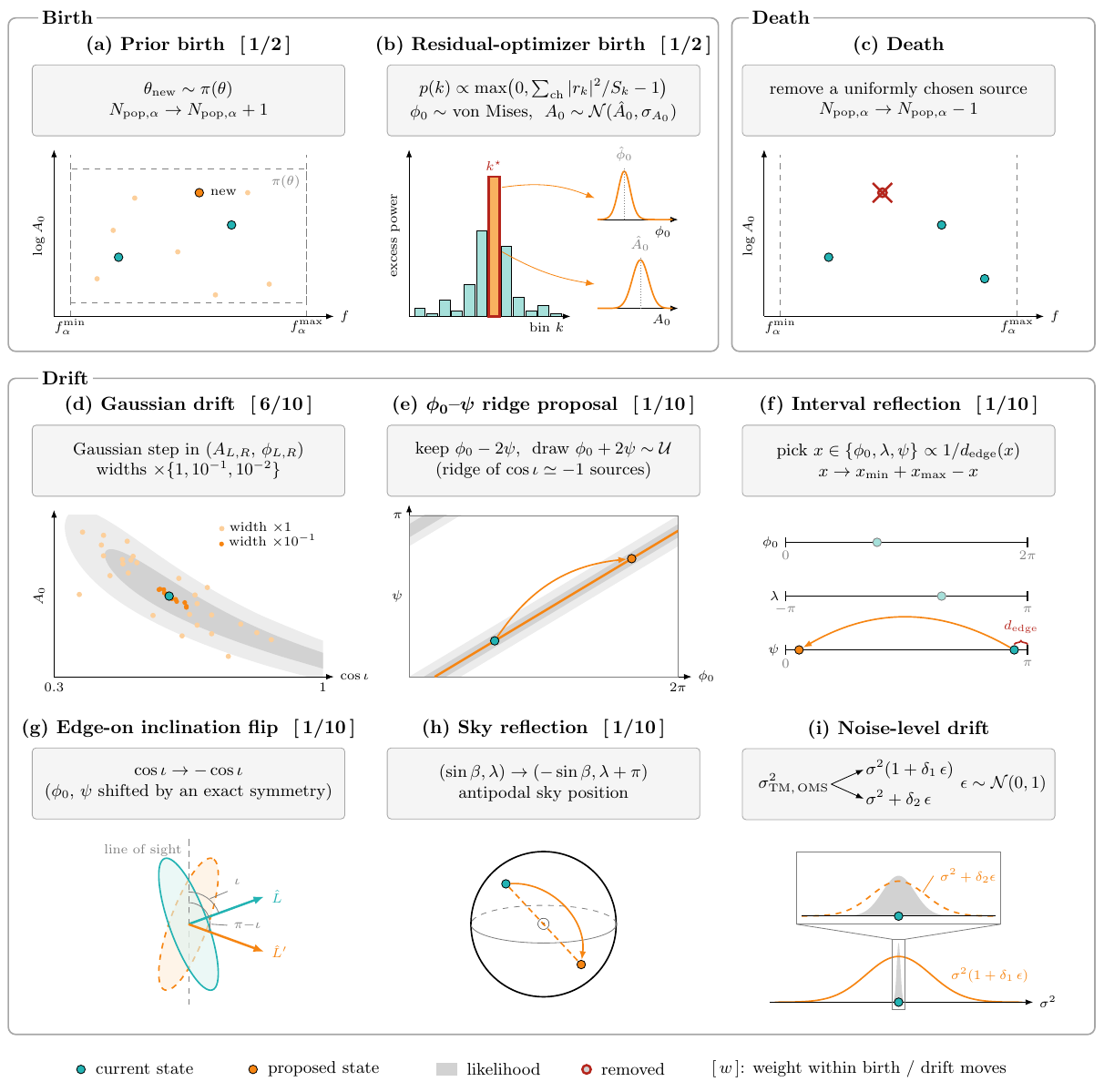}
    \caption{Schematic of the proposals used by \texttt{samsara++} in the mock GF of Sections~\ref{sec:SingleBin},~\ref{sec:Results}, and~\ref{sec:Stress_test}, described in Appendix~\ref{app:proposals}. Teal points mark the current state, orange points the proposed state, and grey shading the likelihood. The bracketed number next to each title is the fraction of birth (mutation) proposals drawn from that sampler. In panels (a)--(c), $N_{\rm pop,\alpha}$ is the number of DWDs in the frequency bin $\alpha$, and $[f_\alpha^{\min}, f_\alpha^{\max}]$ its bounds. In panel (i), the inset zooms into the narrow posterior on the noise level, showing that the additive step $\sigma^2+\delta_2\epsilon$ probes much smaller scales than the relative step $\sigma^2(1+\delta_1\epsilon)$.}
    \label{fig:proposals}
\end{figure*}

In this appendix, we describe the proposals used by \texttt{samsara++} in the mock GFs of Sections~\ref{sec:SingleBin},~\ref{sec:Results}, and~\ref{sec:Stress_test}, illustrated schematically in Figure~\ref{fig:proposals}. Birth and mutation proposals are drawn from several samplers, with relative weights that are also reported in Figure~\ref{fig:proposals}.

For births, we use a mixture of two proposals of equal weight. The former is simply a new point extracted from the prior, which is useful to explore the full parameter space and to ensure a finite $h_b$ in Eqs.~\eqref{eq:birth_death_rates},~\eqref{eq:birth_death_rates_fixed}, and so a non null death rate. The latter is a residual-optimizer birth sampler (ROBS) that spawns sources where the residuals require them. ROBS draws a Fourier bin with probability proportional to the excess power of the residuals over the noise PSD, summed over the AET channels, sampling a frequency uniformly within it. Then, given a unit-amplitude template $g$ with $\phi_0=0$ and the other parameters extracted from the prior, the initial phase is drawn from a von Mises distribution centered on $\arg\langle g|r\rangle$, with concentration set by the matched-filter SNR. The amplitude $A_0$ is drawn from a Gaussian, truncated to the prior range, centered on the amplitude of the matched-filter at the drawn phase. Here, $\langle a|b\rangle \equiv \sum_{f,I} a_I^*(f)\, b_I(f)/S_I(f,\Theta_{\rm n})$ is the complex noise-weighted inner product, with $S_I$ defined below Eq.~\eqref{eq:Whittle}, whose real part enters the likelihood.

The drift proposals act on single DWDs or on the noise parameters. For instrumental noise, the $\sigma^2_{\rm TM}\equiv r_{\rm acc}$ and $\sigma^2_{\rm OMS}\equiv r_{\rm oms}$ of Eq.~\eqref{eq:noise_amplitude_parametrization} are updated with a mixture of two Gaussian steps, one with a relative step with $\delta_1=0.1$ and another one with an additive step with $\delta_2=10^{-4}$ or $10^{-5}$. For DWDs, we use Gaussian steps along some specific combination of the parameters to follow degeneracy curves. The first Gaussian step involves the left- and right-handed amplitudes and phases, $A_{L,R}=\tfrac{1}{4}A_0(1\pm\cos\iota)^2$ and $\phi_{L,R}=\tfrac{1}{2}(\phi_0\pm2\psi)$, which helps the mixing between the unrotated parameters. A second proposal keeps the combination $\phi_0-2\psi$ fixed and draws $\phi_0+2\psi$ uniformly within the prior, moving the source along the likelihood ridge of sources with $\cos\iota\simeq-1$. A third proposal chooses one of the periodic parameters $\phi_0$, $\lambda$, $\psi$ with probability inversely proportional to its distance $d_{\rm edge}$ from the nearest prior boundary, and reflects it about the midpoint of its prior range. A fourth proposal flips the sign of $\cos\iota$, reversing the handedness of the orbit, together with the shifts $\phi_0\to\phi_0+\pi$ and $\psi\to\psi+\pi/2$, which leave the waveform invariant; this move is mostly accepted for nearly edge-on sources. A fifth proposal maps the sky position to its antipode.

For death moves, we adopt a pure death proposal, which simply removes one individual from the society. A crucial point of the \texttt{samsara} approach is that at each step of the CT-MCMC we propose a pure death for each individual in the society. 

\section{Sampler validation}
\label{app:ppplot}

\begin{figure}[h]
    \centering
    \includegraphics[width=0.82\linewidth]{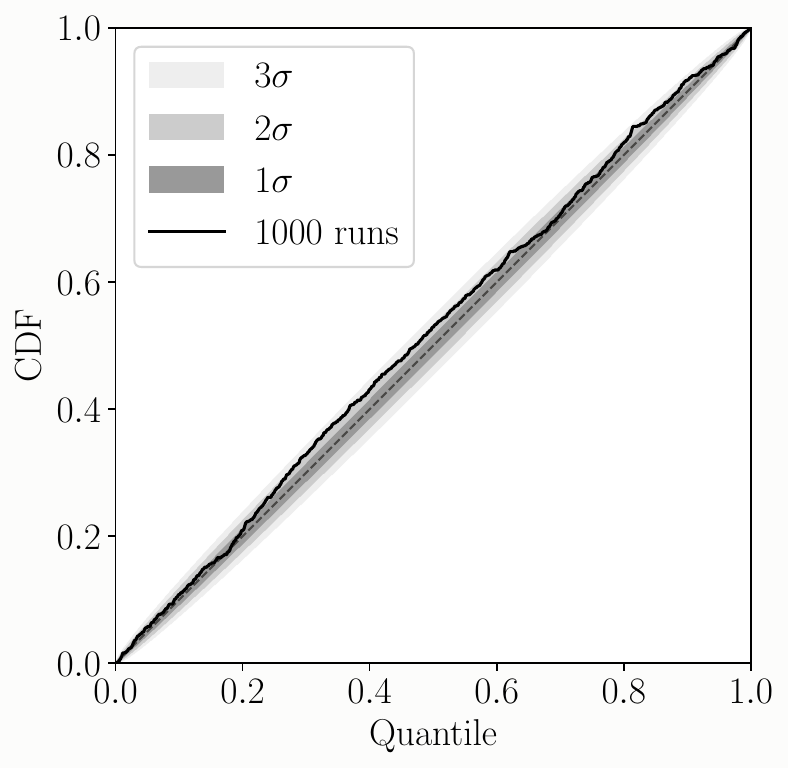}
    \caption{P--P plot for 1000 injections analysed with \texttt{samsara++} using drift moves only. The black line shows the empirical cumulative distribution of the credible-interval quantiles of the injected parameters. The shaded regions mark the $1\sigma$, $2\sigma$ and $3\sigma$ bands expected for a uniform distribution.}
    \label{fig:ppplot}
\end{figure}

To validate the sampler, we ran 1000 simulations. Each one injects a signal drawn from a reference prior and analyses it with \texttt{samsara++} using only drift moves, with no transdimensional evolution. The reference prior is uniform in each DWD parameter, with the bounds listed in Table~\ref{tab:pp_prior_table}; the same prior is used for the injections and for the analysis. Figure~\ref{fig:ppplot} shows the resulting P--P plot. The injected signals have signal-to-noise ratios with median and 90\% confidence interval $\mathrm{SNR} = 5.58^{+7.93}_{-4.33}$. To check that the credible-interval quantiles of the injected parameters follow a uniform distribution, we run two tests. A $\chi^2$ goodness-of-fit test with 30 bins gives a $p$-value of $0.6947$. A Kolmogorov--Smirnov test on the maximum distance between the empirical and expected cumulative distributions gives a $p$-value of $0.1355$. Neither test shows a significant departure from uniformity, so the posteriors recovered by \texttt{samsara++} are consistent with being well calibrated.

\begin{table}[h]
    \centering
    \begin{tabular}{c c}
    \hline
     Parameter & Prior \\
     \hline
     $f_{\rm GW}\, [\rm mHz]$ & $\mathcal{U}(0.99955580, 1.00082420)$ \\
     $\dot{f}_{\rm GW}\, [\rm s^{-2}]$ & $\mathcal{U}(10^{-26}, 10^{-24})$ \\
     $A_0$ & $\mathcal{U}(1.162\times10^{-23}, 1.4\times10^{-22})$ \\
     $\sin \beta$ & $\mathcal{U}(-0.43497, 0.43497)$ \\
     $\lambda$ & $\mathcal{U}(-0.3, 0.2)$ \\
     $\cos \iota$ & $\mathcal{U}(-1, 1)$ \\
     $\psi$ & $\mathcal{U}(0, \pi)$ \\
     $\phi_0$ & $\mathcal{U}(0, 2\pi)$ \\
     \hline
     \hline
    \end{tabular}
    \caption{Reference prior used to draw and analyse the injections used for the P--P plot in Figure~\ref{fig:ppplot}. The bounds on $\sin\beta$ correspond to $|\beta| \leq 0.45\,\rm rad$.}
    \label{tab:pp_prior_table}
\end{table}

\section{Convergence and mixing of the chains}
\label{app:convergence and mixing}

In Section~\ref{sec:SingleBin}, we present the inference of DWDs in a single frequency bin. Here we discuss the specifics of PT and the prior choice to improve the convergence and mixing of the chains.

\begin{figure}
    \centering
    \includegraphics[width=\columnwidth]{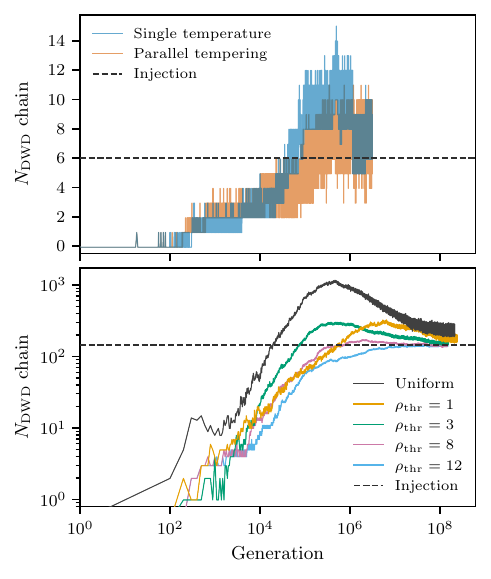}
    \caption{Trace of the number of DWDs as a function of the generation. Top: 6 injected DWDs, for a run with a single temperature and a run with PT with 8 logarithmically spaced temperatures between $T_{\rm min}=1$ and $T_{\rm max}=15$. Bottom: full bin with 143 injected DWDs, without PT, for a uniform prior on the number and for the prior of Eq.~\eqref{eq:general_aic_prior} with $\rho_{\rm thr}=1$, 3, 8, and 12. The dark grey dashed lines mark the number of injected DWDs.}
    \label{fig:PTvs1T_and_prior}
\end{figure}

We note that PT is typically essential for the sampler to enter faster into the main mode of the posterior, especially for $N_{\rm pop}\geq 8$, due to the correlation between the source and the large prior bounds. To show explicitly this, in the top panel of Figure~\ref{fig:PTvs1T_and_prior}, we plot the trace of the number with 6 injections obtained from a run without PT and another with 8 logarithmically spaced temperatures between $T_{\rm min}=1$ and $T_{\rm max}=15$. From the picture it is clear that the chain using PT converges to the number of injected sources in about $10^5$ iterations, whereas the chain with a single temperature slightly overfit signals until it converges around $10^6$: the PT proves to be an order of magnitude more efficient even in this simple case. We stress that, due to the efficient parallelization scheme, the time complexity for the two inferences is comparable because, for the same number of cores, PT exploits at maximum the parallelization.

In the bottom panel of Figure~\ref{fig:PTvs1T_and_prior}, we plot the trace of the number of sources as a function of the generation for three different prior choices. We consider $\rho_{\rm thr}$ equal to 0, which corresponds to a uniform prior, 3, which implies a small penalty in fitting instrumental noise with DWDs, and 12, which produces a large penalty in fitting instrumental noise with DWDs. In order to show only the effect of the prior, we do not rely on PT in this analysis. We notice that during the burn-in phase the chain with the uniform prior overestimates (due to large-SNR sources) by an order of magnitude the number of injections, while the chain with the large-penalty prior underestimates the number of sources asymptotically. We choose $\rho_{\rm thr}=3$ in further analyses, because it is the best compromise between accepting new sources during the CT-MCMC and avoiding too many sources during the burn-in.

